\documentclass[aps,preprint,preprintnumbers,amsmath,amssymb,amsfonts,groupedaddress,superscriptaddress]{revtex4-2}

\usepackage{graphicx}
\usepackage{subfigure}
\usepackage{mathtools}
\usepackage{bm}
\usepackage{color}
\usepackage{xcolor}
\usepackage{array}
\usepackage{cases}
\usepackage{upgreek}

\newcommand{\bmath}{\begin{mathletters}}
\newcommand{\emath}{\end{mathletters}}
\newcommand{\be}{\begin{eqnarray}}
\newcommand{\ee}{\end{eqnarray}}
\newcommand{\ba}{\begin{array}}
\newcommand{\ea}{\end{array}}

\newcommand{\no}{\nonumber}

\newcommand{\h}{\hbar}
\newcommand{\pr}{\prime}

\newcommand{\calT} {\mathcal T}

\begin{document}
\title{Electronic Band Structure of Silicon Determined via a \\Variational Adiabatic Eigensolver: Theory and Experiment }

\author{Xingrui Liu}
\thanks{These authors contributed equally to this work.}
\affiliation{Zhejiang Province Key Laboratory of Quantum Technology and Device, School of Physics, Zhejiang University, Hangzhou 310027, China}
\author{Liyang Sui}
\thanks{These authors contributed equally to this work.}
\affiliation{Zhejiang Province Key Laboratory of Quantum Technology and Device, School of Physics, Zhejiang University, Hangzhou 310027, China}

\author{Tianqi Cai}
\affiliation{Tencent Quantum Laboratory, Tencent, Shenzhen 518057, China}
\author{Zhiwen Zong}
\affiliation{Tencent Quantum Laboratory, Tencent, Shenzhen 518057, China}
\author{Kunliang Bu}
\affiliation{Tencent Quantum Laboratory, Tencent, Shenzhen 518057, China}

\author{Wenyan Jin}
\affiliation{Zhejiang Province Key Laboratory of Quantum Technology and Device, School of Physics, Zhejiang University, Hangzhou 310027, China}
\author{Bowen Chen}
\affiliation{Zhejiang Province Key Laboratory of Quantum Technology and Device, School of Physics, Zhejiang University, Hangzhou 310027, China}
\author{Xutao Zhang}
\affiliation{Zhejiang Province Key Laboratory of Quantum Technology and Device, School of Physics, Zhejiang University, Hangzhou 310027, China}
\author{Yufan Li}
\affiliation{Zhejiang Province Key Laboratory of Quantum Technology and Device, School of Physics, Zhejiang University, Hangzhou 310027, China}

\author{Zhihao Gong}
\affiliation{School of Chemistry, Tianjin Normal University, Tianjin 300387, China}
\author{Yicong Zheng}
\affiliation{Tencent Quantum Laboratory, Tencent, Shenzhen 518057, China}

\author{Shengyu Zhang}
\affiliation{Tencent Quantum Laboratory, Tencent, Shenzhen 518057, China}

\author{Jianlan Wu} \thanks{jianlanwu@zju.edu.cn}
\affiliation{Zhejiang Province Key Laboratory of Quantum Technology and Device, School of Physics, Zhejiang University, Hangzhou 310027, China}

\author{Yi Yin} \thanks{yiyin@zju.edu.cn}
\affiliation{Zhejiang Province Key Laboratory of Quantum Technology and Device, School of Physics, Zhejiang University, Hangzhou 310027, China}
\date{April 15, 2026}

\begin{abstract}
This work addresses the critical challenge of excited-state
preparation for semiconductor band structure calculations.
We introduce a variational adiabatic eigensolver (VAE) protocol
that combines adiabatic evolution with variational optimization
to prepare high-fidelity eigenstates on noisy intermediate-scale
quantum (NISQ) devices. Applying a momentum-space truncation,
we accurately compute the electronic band structure of silicon---an
idealized infinite periodic system---using only a modest number of
qubits. Our approach employs multi-qubit parameterized circuits
and a phase-based loss function, overcoming limitations of
conventional methods. These limitations include the circuit-construction
difficulty in traditional adiabatic approaches and the reduced
accuracy of variational quantum eigensolvers for excited states.
Through rigorous numerical simulation and experimental implementation
on a superconducting quantum processor, we successfully prepare
silicon\textquoteright s valence-band and conduction-band eigenstates.
Single-shot readout yields state fidelities exceeding 96\%,
and the measured energy expectations agree with theoretical band
energies within 0.5 eV. Further refinement via single-frequency
oscillation fitting reduces the energy deviation to below 0.01 eV.
This framework provides a robust and practical pathway for precisely
determining electronic structures in quantum materials.
\end{abstract}

\maketitle

\newpage
\section{Introduction}
\label{sec1}

Quantum computation has emerged as a rapidly advancing field, fueled by significant progress
in both hardware development and algorithmic
innovations~\cite{Nielsen2010book,AruteNature19,AcharyaNature25}.
It demonstrates the capability to address a broad spectrum of fundamental physical problems
across diverse quantum platforms, including simulating condensed matter
systems~\cite{AruteScience20},
determining electronic structures in quantum
chemistry~\cite{McArdleRMP20,LanyonNatureChem10}, and exploring
quantum thermodynamics~\cite{VinjanampathyContempPhys16}.
However, inherent errors in quantum manipulation and measurement constrain current devices
to the noisy intermediate-scale quantum (NISQ) era~\cite{Preskill18},
limiting their practical applications. This underscores the need for robust algorithms
capable of operating under such constraints.

The determination of microscopic electronic structures governs the fundamental properties
of materials and molecules in quantum physics and chemistry. Classical computational
methods rely on Hamiltonian diagonalization, a process that faces exponential scaling
with system size, creating a fundamental bottleneck for studying large systems.
In contrast, quantum algorithms leverage the principles of quantum state evolution to
overcome this limitation. For instance, early
quantum phase estimation (QPE) algorithms~\cite{Nielsen2010book,LanyonNatureChem10,AspuruGuzikScience05}
extract eigenenergies from the phase evolution of a system's eigenstate.
However, QPE requires complex circuits involving auxiliary qubits and quantum
Fourier transforms, making it highly sensitive to noise and demanding extremely high
gate fidelities that are challenging to achieve on current hardware.
Variational methods~\cite{CerezoRev2021}, such as the variational quantum
eigensolver (VQE)~\cite{PeruzzoNatCommun14,McCleanNewJPhys16},
offer a more practical alternative for NISQ devices.
VQE employs parameterized circuits optimized via hybrid quantum-classical workflows
to estimate ground states.

A number of extended VQE schemes have been developed for
excited-state energy calculations~\cite{TillyPhysRep22},
including variational quantum deflation (VQD)~\cite{HiggottQuantum19},
automatically-constrained VQE (VQE/AC)~\cite{Gocho2023},
variance-minimization VQE~\cite{Zhang2022,Zhang2021},
the folded spectrum method~\cite{Tazi2024},
and subspace-search VQE (ss-VQE)~\cite{Nakanishi2019}.
Despite their theoretical advances, these methods face distinct
practical limitations~\cite{TillyPhysRep22,Xie2022}: VQD and VQE/AC require precomputation and
explicit orthogonalization against lower-energy states, leading to error accumulation;
variance-based approaches have difficulties in target-state convergence;
the folded spectrum method depends on accurate initial excited-state estimates;
while ss-VQE demands extensive parameterization and deep circuits for high-energy states.

Apart from the variational method, adiabatic evolution is
another approach for preparing eigenstates~\cite{Farhi2000}.
Its fundamental principle is to evolve the system
along a slowly varying Hamiltonian path from an easily
prepared initial state to a final state that
encodes the solution~\cite{Farhi2000},
enabling access to both ground and excited states without
inherent distinction. These algorithms can be broadly categorized
into analog and digitized implementations. Analog
adiabatic eigensolvers (AAE) utilize
continuous, hardware-native control of Hamiltonian parameters over time.
Quantum annealing serves as a prominent example, operating through
continuous evolution on specialized hardware to locate ground states
of combinatorial optimization problems~\cite{JohnsonNature11}. AAE are limited
to hardware-compatible Hamiltonians, restricting their programmability.
The digitized adiabatic eigensolver (DAE) provides a more general
and programmable alternative. By leveraging Trotter-Suzuki
decomposition~\cite{lloyd96}, DAEs approximate the continuous adiabatic
evolution as a finite sequence of parameterized quantum gates---enabling
execution on universal, gate-based quantum
processors~\cite{Lamata2018,Salathe2015,BarendsNature16}.
However, the depth of the DAE circuit is limited by the number of Trotter
steps which directly proportional to evolution time, leading
to error accumulation on noisy devices. It is essential to
refine adiabatic quantum algorithms
to ensure their practicality and robustness in different applications.

In this work, we introduce a variational adiabatic eigensolver (VAE),
a hybrid framework that integrates the state-preparation guarantees
of adiabatic evolution with the flexibility and NISQ-compatibility
of variational optimization. This approach is specifically designed to
overcome the persistent challenges in excited-state preparation.
Our method employs hardware-efficient~\cite{KandalaNature17}, multi-qubit parameterized circuits
and introduces a phase-based loss function to explore high-energy
excited eigenstates. The VAE protocol operates by constructing an adiabatic path where,
at each step, a parameterized circuit is variationally optimized to
approximate the corresponding eigenstate. This approach effectively
replaces the deep quantum circuits required for digitized adiabatic
evolution with a sequence of shallow, optimized circuits.

While our VAE is developed following our
earlier DAE work~\cite{Zhan2021},
we also find that similar adiabatic-variational principles have been
explored in prior
literature~\cite{GarciaSaezLatorre18,Matsuura2020,Matsuura2021,Harwood2022,Schiffer2022}.
Previous studies focused predominantly on ground-state problems,
where conventional variational methods often succeed without rigorous
adiabatic guidance---leaving the broader utility of such pathways underexplored.
Our approach targets excited-state computation, where adiabatic
constraints play an essential role: they enforce continuous evolution
along the desired excited-state, avoiding leakage into
unintended eigenstates---a possible failure of other variational
optimization for excited states. Adiabatic evolution is also particularly
well-suited for calculating electronic band structures. By following
a continuous path in the Brillouin zone, the eigenstates and eigenenergies
at intermediate momenta emerge as natural byproducts, thereby
simplifying the determination of the full band structure.

We validate the VAE framework through comprehensive numerical
simulations and experimental implementation on a superconducting
quantum processor~\cite{BarendsPRL13,Krantz19}. The silicon band structure is selected as
a test case due to its dual role as a realistic solid-state
system and a well-established benchmark in electronic
structure theory. Our protocol
begins by preparing the system in an easily initializable
eigenstate of a simple Hamiltonian $H_0$. Using an adiabatic
evolution process, the system is guided to a pre-selected anchor
point---typically a high-symmetry wavevector in the Brillouin
zone---to prepare the corresponding Bloch eigenstate.
This anchor state then serves as the starting point for a
sequence of sequential adiabatic steps along the band structure,
where each subsequent wavevector $\vec{k}$  is treated as an
intermediate target. In this manner, eigenstates at
different $\vec{k}$-points are computed efficiently along
a well-defined adiabatic trajectory.

\begin{figure}[tp]
    \centering
    \includegraphics[width=0.75\linewidth]{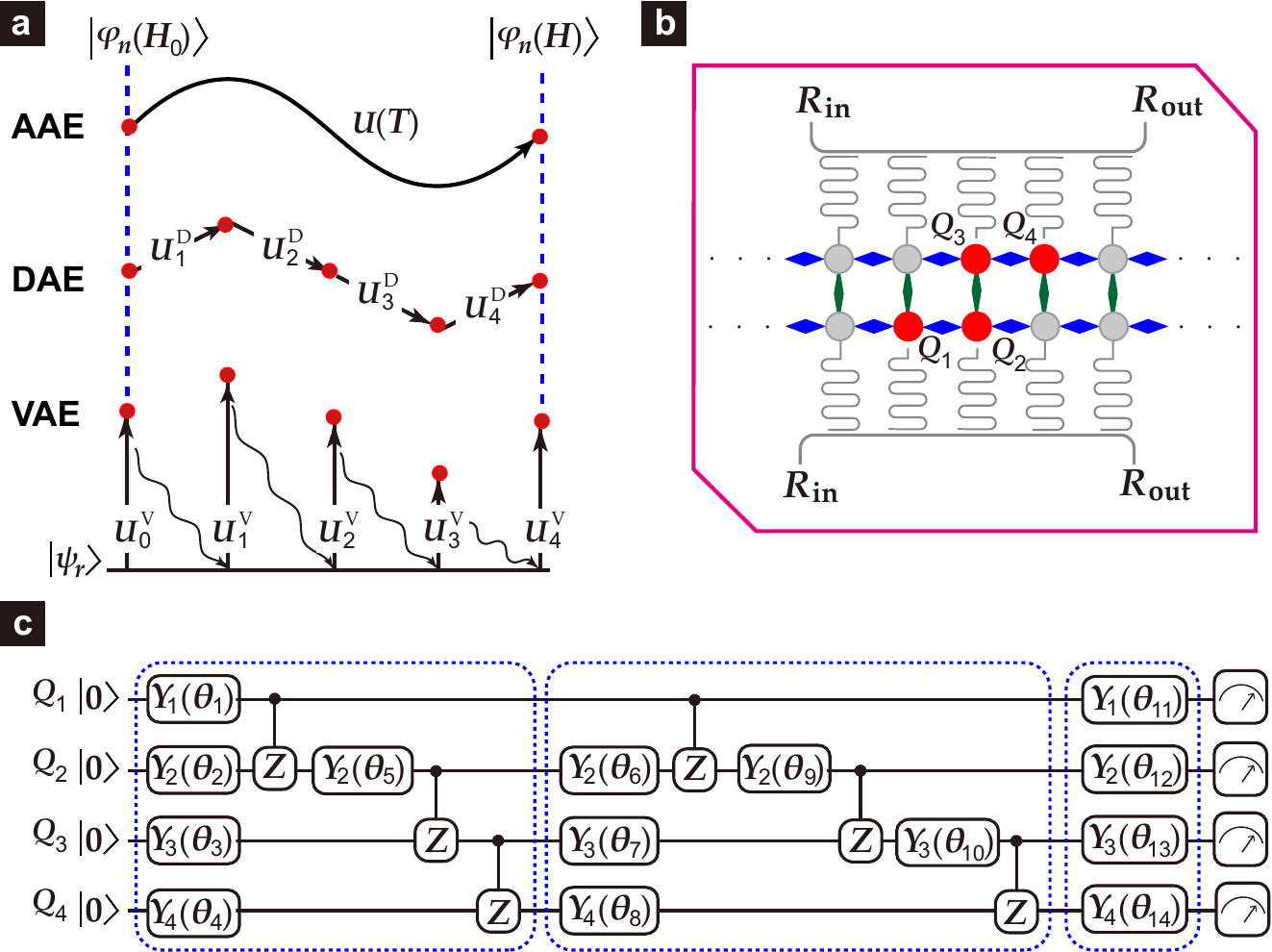}
    \caption{(a) The schematic diagrams of three adiabatic eigensolvers:
    analog (AAE), digitized (DAE) and variational (VAE),
    which realize the same evolution from the initial eigenstate
    $|\varphi_{n}(H_0)\rangle$ to the target eigenstate $|\varphi_n(H)\rangle$
    via three different algorithms.
    (b) A selected portion of a superconducting quantum processor
    with a dual-row 40-qubit architecture. The 4 working qubits are highlighted
    in red color while the others are in grey color. The intra-row couplers (blue) are tunable
    while the inter-row couplers (green) are fixed.
    Each qubit is connected to a readout line through an independent resonator,
    and each readout line integrates five qubits and their resonators.
    (c) The 2.5-layer ansatz quantum circuit in our VAE computation of the silicon band structure.
    Each layer includes a number of single-qubit $Y$ gates
    and two-qubit C$Z$ gates. The 14 rotational angles $\theta_l$ ($l = 1, 2, \dots, 14$)
    are determined through variational optimization.
    }
    \label{fig1}
\end{figure}

By combining a phase-based loss function with customized
adiabatic paths, our approach supports controlled
state evolution along high-symmetry directions in $k$-space.
It enables high-fidelity preparation
of high-energy excited states in crystalline silicon---such as
the fourth and fifth excited states or the valence band
and conduction band.
Furthermore, the method is designed to handle realistic band
structure complexities including energy-level crossings,
anti-crossings, and degenerate states, going beyond the
simplified assumption of isolated eigenstates.

Experimentally, we demonstrate high-fidelity preparation of
the valence and conduction band states of crystalline
silicon---with measured
fidelities exceeding 96\%. The band energies are extracted
directly from the experimentally prepared states
by evaluating the expectation value of the system Hamiltonian,
with a deviation smaller than 0.5 eV.
Additionally, we employ a single-frequency fitting procedure
on the expectation of the phase-based time evolution operator,
yielding band energies within 0.01 eV of theoretical
values. This work establishes VAE as a practical
and reliable tool for simulating solid-state bands
on current NISQ devices, highlighting its potential
for enabling robust electronic structure calculations
of complex quantum materials.

\section{Hamiltonian of Silicon}
\label{sec2}

Silicon crystallizes in the diamond cubic structure, formed by
displacing two interpenetrating face-centered cubic sublattices
along the body diagonal by one quarter of its length~\cite{Hamaguchi2010book}.
The corresponding
unit cell can be viewed as a tetrahedron containing two silicon
atoms, positioned at $\vec{R}_1\mspace{-3mu}=\mspace{-3mu}(a/8, a/8, a/8)$
and $\vec{R}_2 \mspace{-3mu}=\mspace{-3mu} -\vec{R}_1$, where
$a \mspace{-3mu}=\mspace{-3mu}5.43~\mathrm{\AA}$ is the lattice
constant~\cite{Hamaguchi2010book}. In order to obtain a real-valued
Hamiltonian matrix for quantum computation, the origin of the
Bravais lattice is shifted to the midpoint $(\vec{R}_1 + \vec{R}_2)/2$.

The electronic band structure of silicon is typically computed using sophisticated methods
like density functional theory (DFT)~\cite{Hohenberg1964,Kohn1965,Martin2004}.
In this work, however, we adopt the nearly free electron (NFE) model~\cite{Kittel2005}
with an empirical pseudopotential~\cite{Martin2004,Hamaguchi2010book}
for computational efficiency and simplicity. The pseudopotential
for each valence electron is expressed as
$ V(\vec{r})\mspace{-3mu}=\mspace{-3mu}\sum_{n, i} v(\vec{r}\mspace{-3mu}-\mspace{-3mu}\vec{R}_{n, i})$,
where $v(\vec{r}\mspace{-3mu}-\mspace{-3mu}\vec{R}_{n, i})$
denotes a smoothed pseudopotential
centered at  the position $\vec{R}_{n, i=1, 2}$ of the $i$-th silicon
atom in the $n$th unit cell. Due to the translational
periodicity of the crystal, the pseudopotential
is expanded in terms of reciprocal lattice vectors (RLVs) $\vec{G}$ as
$V(\vec{r})\mspace{-3mu}=\mspace{-3mu}\sum_{\vec{G}} V_{\vec{G}}\exp(i\vec{G}\mspace{-1mu}\cdot\mspace{-1mu}\vec{r})$,
with the Fourier coefficient given by $V_{\vec{G}}\mspace{-3mu}=\mspace{-3mu}\Omega^{-1} S_{\vec{G}}v_{\vec{G}}$.
Here, $\Omega=a^3/4$ is the unit cell volume,
$S_{\vec{G}} \mspace{-3mu}=\mspace{-3mu} 2\cos(\vec{G}\mspace{-2mu}\cdot\mspace{-2mu}\vec{R}_1)$ is
the structure factor, and $v_{\vec{G}}\mspace{-3mu}\approx\mspace{-3mu}\int_{\mathrm{cell}}v(\vec{r})
\exp(-i\vec{G}\mspace{-2mu}\cdot\mspace{-2mu}\vec{r})d\vec{r}$
represents the Fourier transform of the local pseudopotential $v(\vec{r})$.

Within the first Brillouin zone (FBZ), a Bloch state corresponding to wavevector
$\vec{k}$ is expanded as
$|\varphi_{\vec{k};\vec{G}}\rangle\mspace{-3mu}=\mspace{-3mu}\sum_{\vec{G^\pr}}C_{\vec{G},\vec{G}^\pr}
\exp[i(\vec{k}+\vec{G}^\pr)\mspace{-2mu}\cdot\mspace{-2mu}\vec{r}\mspace{1.5mu}]$,
where $\vec{G}$ indexes the electronic band.
Substituting into the Schr\"{o}dinger equation yields the central equation~\cite{Kittel2005}:
\be
(\lambda_{\vec{k}+\vec{G}}-\mathcal{E}_{\vec{k};\vec{G}})C_{\vec{G},\vec{G}}+\sum_{\vec{G}^\pr} V_{\vec{G}-\vec{G}^\pr}C_{\vec{G}^\pr, \vec{G}}=0,
\label{eq2_1}
\ee
in which $\lambda_{\vec{k}+\vec{G}}=\hbar^2(\vec{k}+\vec{G})^2/2m$
denotes the free-electron kinetic energy.
Equation~(\ref{eq2_1}) can also be written in a matrix form as
\be
HC = \mathcal{E} C,
\label{eq2_2}
\ee
where the effective Hamiltonian matrix elements are
defined as $H_{\vec{G},\vec{G}^\pr}=\lambda_{\vec{k}+\vec{G}}\delta_{\vec{G}, \vec{G}^\pr}-V_{\vec{G}-\vec{G}^\pr}$.

Assuming spherical symmetry ($v_{\vec{G}} \mspace{-3mu}=\mspace{-3mu} v_{G}$)
and with the condition of $G\le 2b$ ($b\mspace{-3mu}=\mspace{-3mu}2\pi/a$),
components beyond $\Delta G>\sqrt{11}b$ are neglected and only lowest
15 RLVs are selected. Correspondingly ,
we adopt the following empirical values
$v_{G=0}\mspace{-3mu}=\mspace{-3mu}-5.66$~eV,
$v_{G=\sqrt{3}b}\mspace{-3mu}=\mspace{-3mu}-1.43$~eV, $v_{G=\sqrt{8}b}\mspace{-3mu}=\mspace{-3mu}0.27$~eV,
$v_{G=\sqrt{11}b}\mspace{-3mu}=\mspace{-3mu}0.54$~eV,
as given in Ref.~\cite{Hamaguchi2010book,Brust1964}.
Diagonalization of the Hamiltonian constructed from
these 15 RLVs yields the band structure
$\mathcal{E}=\mathcal{E}_{\vec{k};\vec{G}}$.
The results have close agreement with independent DFT calculations,
confirming the validity of the truncated NFE model for silicon
in our quantum computational approach.

To implement quantum eigensolver algorithms on a quantum
processor~\cite{AspuruGuzikScience05,PeruzzoNatCommun14,BarendsNature16,KandalaNature17,Zhan2021,CollessPRX18},
the effective Hamiltonian $H$ must be transformed into a representation
using Pauli operators. This transformation enables the Hamiltonian
to be executed on qubit-based quantum hardware. Following the
principles of second quantization, the creation and annihilation
operators of electrons are mapped to products of Pauli operators
through established techniques such as the Jordan-Wigner transformation
and the Bravyi-Kitaev transformation~\cite{Whitfield2011,BravyiKitaev2002,Seeley2012,Tranter2015}.
In our specific case, diagonalizing the Hamiltonian $H$ over 15 RLVs
would nominally require a 15-qubit quantum processor.
However, under the single-electron approximation inherent in
the NFE model, the effective Hamiltonian operates within a
15-dimensional Hilbert space, which is considerably smaller
than the full $2^{15}$-dimensional space of 15 qubits.
To optimize resource utilization, we therefore select a
4-qubit system, corresponding to a $2^4$-dimensional Hilbert space.
To accommodate the 15-dimensional electronic structure
within this 4-qubit framework, the Hamiltonian $H$
is expanded to include one auxiliary dummy state, thereby
fulfilling the $2^4$-dimensional space requirement.
While we have reduced the qubit number, simulating dense (as opposed
to sparse) Hamiltonians on multi-qubit processors poses
some experimental challenges.
An electronic state in this 4-qubit system is expressed
as $|\psi\rangle \mspace{-3mu}=\mspace{-3mu} \sum_S c_S |S\rangle$,
where $|S\rangle \mspace{-3mu}=\mspace{-3mu} |s_1 s_2 s_3 s_4\rangle$
represents a computational basis state. For each qubit
$i$ ($=\mspace{-3mu}1,2,3,4$), $|s_i\rangle$ denotes
either the ground state or excited state,
i.e., $|s_i\rangle\mspace{-3mu} \in \mspace{-3mu}\{|0\rangle_i, |1\rangle_i\}$.
We then introduce an operator set
$\{\mathcal{O}_\alpha\}\mspace{-3mu}=\mspace{-3mu} \{\sigma_1\sigma_2\sigma_3\sigma_4 \}$
containing $M(=\mspace{-3mu}4^4)$ elements, where each
$\sigma_i$ represents a Pauli operator from the set
$\{I, X_i, Y_i, Z_i\}$.
Using this operator basis, the effective Hamiltonian in Eq.~(\ref{eq2_2})
is transformed into the Pauli representation~~\cite{Koska2024}:
\be
H=\sum_{\alpha=1}^M h_\alpha \mathcal{O}_\alpha
\label{eq2_3}
\ee
where the expansion coefficients are given by
$h_\alpha\mspace{-3mu}=\mspace{-3mu} \mathrm{tr}\{ H \mathcal O_\alpha\}$.
This formulation enables the implementation of quantum eigensolver algorithms
on quantum hardware.

Conventional VQE is typically applied to small molecules in real space,
where both the system size and the number of required qubits are limited.
In contrast, we simulate an infinite silicon crystal using only a small
number of qubits---a feat made possible by truncating the problem in
momentum space rather than real space. We start with 15 RLVs, a basis that
can be systematically increased to 113 RLVs to improve
convergence~\cite{Hamaguchi2010book,Brust1964}. Although the resulting
15-dimensional Hamiltonian is easily diagonalized classically, we
intentionally choose this minimal case as a clear and pedagogical benchmark,
analogous to the hydrogen molecule in conventional VQE benchmarks.
Within this NFE framework, we thereby demonstrate the implementation and
efficacy of our adiabatic quantum algorithm for electronic
structure calculations.

\section{Variational Adiabatic Eigensolver}
\label{sec3}

The adiabatic path in our VAE method is adapted from the AAE
framework~\cite{Farhi2000},
which pioneered some applications in quantum computing.
Based on the adiabatic theorem,
the AAE employs a time-dependent Hamiltonian
$H(t)\mspace{-3mu}=\mspace{-3mu}H_0\mspace{-3mu}+\mspace{-3mu}\lambda(t)(H\mspace{-3mu}-\mspace{-3mu}H_0)$,
where coefficient $\lambda(t)$  evolves smoothly from 0 to 1 over duration $T$.
When the system is initialized in an eigenstate of $H_0$ and evolves sufficiently
slowly ($\dot{\lambda}(t)\rightarrow 0$), it remains in the
corresponding eigenstate of $H(t)$, yielding the target eigenstate
$|\psi(t\mspace{-3mu}=\mspace{-3mu}T)\rangle\mspace{-3mu}\sim\mspace{-3mu}|\varphi_n(H)\rangle$
and energy $\mathcal E_n \mspace{-3mu}=\mspace{-3mu} \langle \psi(T)|H|\psi(T)\rangle$ [Fig.~\ref{fig1}(a)].
For Hamiltonians expressed as $H(t)\mspace{-3mu}=\mspace{-3mu}\sum_{\alpha}h_\alpha(t) \mathcal{O}_\alpha$
in the Pauli basis, the time-dependent coefficients $\{h_\alpha(t)\}$
can be implemented via microwave control~\cite{Krantz19}.
A shortcut-to-adiabaticity (STA) technique has also been
developed to accelerate the adiabatic process in widespread
applications~\cite{Demirplak03,Berry09,ChenXi10,Campo13,ZhangNJP18,WangNJP18,WangPRApp19}.

DAE overcome the hardware-specific constraints of AAE
by compiling the time-dependent Hamiltonian---expressed in an
arbitrary Pauli basis---into a sequence of quantum gates.
This gate-based implementation requires discretizing the total
evolution time $T$ into $N_t$ segments of duration
$\Delta t\mspace{-3mu}=\mspace{-3mu}T/N_t$ to approximate the
underlying continuous-time adiabatic path.
The full evolution operator is factorized as
$U\mspace{-3mu}=\mspace{-3mu}U^\mathrm{D}_{N_t}\cdots U^\mathrm{D}_{2}U^\mathrm{D}_{1}$,
where each segment $U^\mathrm{D}_j\approx \exp[-(i/\hbar)H_{j-1}\Delta t]$
with $H_{j-1}\mspace{-3mu}=\mspace{-3mu}H(t\mspace{-3mu}=\mspace{-3mu}(j\mspace{-1mu}-\mspace{-1mu}1)\Delta t)$.
Through Trotter-Suzuki decomposition, each $U^\mathrm{D}_j$ is decomposed into
a sequence of quantum gates:
\be
\left\{\ba{ccl} |\varphi_n(H_{j})\rangle &\approx& U^\mathrm{D}_j |\varphi_n(H_{j-1})\rangle  \\
U^\mathrm{D}_j &=& u^{\mathrm{D}}_{j, M_\mathrm{D}}\cdots u^{\mathrm{D}}_{j, 2} u^{\mathrm{D}}_{j, 1} \ea \right .
\label{eq3_1}
\ee
where individual gates
$u^{\mathrm{D}}_{j, l=1,\cdots, M_D}\mspace{-3mu}=\mspace{-3mu}\exp(-i\theta_{j, l}\mathcal O^\mathrm{D}_l/2)$
are characterized by rotation angles $\theta_{j, l}$ and a Pauli operators
$\mathcal O^\mathrm{D}_{l}$.
This digitized approach enables sequential state evolution
$|\varphi_n(H_0)\rangle\mspace{-3mu} \rightarrow \mspace{-3mu}| \varphi_n(H_1)\rangle
\cdots  \mspace{-3mu} \rightarrow  \mspace{-3mu}| \varphi_n(H)\rangle$
[Fig.~\ref{fig1}(a)] while maintaining compatibility with gate-based quantum processors.

While the DAE performs well for systems with simple interactions~\cite{Zhan2021},
it faces considerable difficulties when applied to Hamiltonians
involving complex terms---such as the full matrix in Eq.~(\ref{eq2_3}).
These include deep quantum circuits from Trotterization,
which amplify hardware noise, as well as constraints imposed
by qubit connectivity that limit accurate Hamiltonian
representation.
To address these issues while retaining the adiabatic pathway
structure, we integrate variational optimization with the
adiabatic framework, resulting in a VAE. Rather than implementing
the exact digitized evolution operator $U^{\mathrm{D}}_j$ (transform $|\varphi_n(H_{j-1})\rangle$ to $|\varphi_n(H_j)\rangle$),
we reinterpret each adiabatic step as a variational
state-preparation task. This is realized via a parameterized
ansatz circuit $U^{\mathrm{V}}_{j}$ applied to a
reference state $|\psi_r\rangle$, generating the trial state:
\be
\left\{\ba{ccl} |\psi_{j}\rangle &=& U^\mathrm{V}_{j} |\psi_r\rangle  \\
               U^\mathrm{V}_{j} &=& u^\mathrm{V}_{j, M_\mathrm{V}}
                                     \cdots u^\mathrm{V}_{j, 2}u^\mathrm{V}_{j, 1} \ea \right. .
\label{eq3_2}
\ee
Each quantum gate $u^\mathrm{V}_{j, l=1, \cdots, M_V}\mspace{-3mu}=\mspace{-3mu}\exp(-i\theta_{j, l} \mathcal O^\mathrm{V}_l/2)$
is characterized by a rotational angle $\theta_{j,l}$ and a Pauli operator
$\mathcal O^\mathrm{V}_{l}$.
Crucially, both the parameter set ${\boldsymbol{\theta}}_j\mspace{-3mu}=\mspace{-3mu}\{\theta_{j, l}\}$
and the operator set
$\boldsymbol{\mathcal O}^\mathrm{V}\mspace{-3mu}=\mspace{-3mu}\{\mathcal O^\mathrm{V}_l\}$ differ from their DAE counterparts, as does the total number of gates $M_\mathrm{V}$
in the VAE circuit.

With an appropriate loss function, variational optimization of circuit
parameters can be carried out to obtain the optimal solution,
${\boldsymbol{\theta}}_j\mspace{-3mu}\rightarrow\mspace{-3mu}{\boldsymbol{\theta}}^\mathrm{opt}_j$.
The resulting state  $|\psi_{j}\rangle_{\mathrm{opt}}\mspace{-3mu}=\mspace{-3mu}
U^\mathrm{V}({\boldsymbol{\theta}}^\mathrm{opt}_j)|\psi_r\rangle$
provides the best approximation to the target eigenstate $|\varphi_n(H_j)\rangle$.
The complete VAE protocol proceeds sequentially through the adiabatic path:
$|\psi_0\rangle_{\mathrm{opt}}\mspace{-3mu} \rightarrow \mspace{-3mu}
|\psi_1\rangle_{\mathrm{opt}} \mspace{-3mu}\rightarrow\mspace{-3mu} \cdots
\mspace{-3mu}\rightarrow\mspace{-3mu} |\psi_{N_t}\rangle_{\mathrm{opt}}$
yielding a high-fidelity approximation to the target eigenstate, i.e.,
$|\psi_{N_t}\rangle_{\mathrm{opt}}\mspace{-3mu} \approx\mspace{-3mu} |\varphi_n(H)\rangle$. The schematic diagram of this variational approach
is presented in Fig.~\ref{fig1}(a), illustrating
how the VAE maintains the conceptual structure of adiabatic evolution while
leveraging the flexibility of variational quantum circuits.

Before presenting the numerical results of the VAE
applied to silicon band structure computation, we address several key technical
considerations crucial for its implementation:
(i.) Reference state preparation. The reference state $|\psi_r\rangle$ must be
efficiently preparable on quantum hardware. A natural choice is the ground state
of the multi-qubit processor. (ii.) Ansatz quantum circuit design.
The parameterized quantum circuit must balance expressibility and trainability
to effectively explore the relevant Hilbert space. We adopt a hardware-efficient (HE)
ansatz architecture~\cite{KandalaNature17}, comprising
multiple layers of single-qubit $Y$-rotations and two-qubit controlled-$Z$ (CZ) gates.
This design aligns with native gate sets of superconducting quantum processors
while maintaining sufficient expressiveness for eigenstate preparation.
The specific HE circuit configuration used in our study is illustrated in
Fig.~\ref{fig1}(c).
(iii.) Loss function formulation.
The VAE relies on the optimization of a carefully constructed loss
function $\mathcal F({\boldsymbol{\theta}}_j)$. Conventional VQE
approaches minimize energy expectations---a strategy effective
for ground-state
calculations~\cite{PeruzzoNatCommun14}.
While excited-state extensions such as VQD~\cite{HiggottQuantum19},
variance-VQE~\cite{Zhang2022,Zhang2021}, and VQE/AC~\cite{Gocho2023} have been proposed,
they may face different challenges and suffer from robustness
under realistic noisy conditions.

Here we introduce a phase-based loss function defined as:
\be
\mathcal{F}(\calT; {\boldsymbol{\theta}}_j)=1-\left|\mathcal W(\mathcal T; {\boldsymbol{\theta}}_j)\right|,
\label{eq3_3}
\ee
where $\mathcal W(\calT; {\boldsymbol{\theta}}_j) =
\langle\psi_j({\boldsymbol{\theta}}_j)|\mathcal U_j(\calT)|\psi_j({\boldsymbol{\theta}}_j)\rangle$
is the expectation of the time evolution operator
$\mathcal U_j(\calT)\mspace{-3mu}=\mspace{-3mu}\exp(-i H_j \calT/\hbar)$
and $\calT$  is an empirically chosen evolution time
independent of the adiabatic schedule.
This function achieves its global minimum,
$\mathcal{F}_\mathrm{min}=0$, where $|\psi_j\rangle$ is
an eigenstate of $H_j$,
providing unbiased treatment of all eigenstates.
(iv.) Time evolution implementation. For the time evolution operator
$\mathcal U_j(\calT)$, we employ a 4th-order Runge-Kutta expansion,
i.e., $\mathcal U_j(\calT)\mspace{-3mu}=\mspace{-3mu} [\Delta \mathcal U_j]^{N_\calT}$
with $\Delta\mathcal U_j\mspace{-3mu}=\mspace{-3mu}\sum_{k=0}^4 (-i H_j\Delta\calT/\hbar)^k/k!$,
where $\Delta\calT$ is a small time
step and $N_\calT \mspace{-3mu}=\mspace{-3mu}\calT/\Delta \calT$ is the number of time steps.
(v.) Parameter initialization strategy.
Successful optimization critically depends on initial parameters ${\boldsymbol{\theta}}^{(0)}_j$.
We employ adiabatic continuity by initializing
${\boldsymbol{\theta}}^{(0)}_j\mspace{-3mu}=\mspace{-3mu}{\boldsymbol{\theta}}^\mathrm{opt}_{j-1}$,
leveraging the spectral similarity between adjacent Hamiltonians $H_{j-1}$ and $H_j$~\cite{Zhan2021}.
This ensures convergence to the target eigenstate
$|\psi_{j}\rangle_\mathrm{opt}\mspace{-3mu}\simeq\mspace{-3mu}|\varphi_{n}(H_j)\rangle$,
except near band crossings where special care is needed (discussed in Sec.~\ref{sec4}).
(vi.) Optimization methodology. The parameter update follows a gradient-based scheme:
\be
{\boldsymbol{\theta}}^{\mathrm{opt}}_{j-1} \xrightarrow{\substack{\nabla \mathcal F^{(0)}}}
{\boldsymbol{\theta}}^{(1)}_j \cdots \xrightarrow{\substack{\nabla \mathcal F^{(s)}}} {\boldsymbol{\theta}}^{(s+1)}_j\cdots,
\no
\ee
where $s$ denotes the $s$th iteration step of the $j$th VAE procedure.
Based on the parameter-shift rule~\cite{MitaraiPRA18}, the state
$|\psi_l\rangle\mspace{-3mu}=\mspace{-3mu}U^\mathrm{V}({\boldsymbol{\theta}})|\psi_r\rangle$
is the trial state, and the shifted states
$|\psi^\pm_l\rangle\mspace{-3mu}=\mspace{-3mu}U^\mathrm{V}({\boldsymbol{\theta}}^\pm_l)|\psi_r\rangle$
are generated using parameter vectors
${\boldsymbol{\theta}}^\pm_l\mspace{-3mu}=\mspace{-3mu}\{\cdots, \theta_l\pm\pi/2, \cdots\}$.
The partial derivative with
respect to a single parameter $\theta_l$ is given by:
\be
\frac{\partial \mathcal F}{\partial \theta_l}
=
-\frac{1}{2|\mathcal W|}
\mathrm{Re}\left[\left(\mathcal W^+_l-\mathcal W^-_l\right)\mathcal W^\ast\right],
\label{eq5_1}
\ee
where $\mathcal W_l\mspace{-3mu}=\mspace{-3mu}\langle\psi_l|\mathcal U(\calT)|\psi_l\rangle$
and $\mathcal W^\pm_l\mspace{-3mu}=\mspace{-3mu}\langle\psi^\pm_l|\mathcal U(\calT)|\psi^\pm_l\rangle$.
Estimating a single partial derivative
requires the generation and measurement
of three distinct quantum states. With such gradients, we further apply the
Broyden-Fletcher-Goldfarb-Shanno (BFGS) algorithm~\cite{Nocedal2016book}
to accelerate the converging speed, with the iteration terminated when
the loss function reaches its convergence tolerance.
This comprehensive technical framework enables
robust VAE implementation for band structure calculations in our following results.

\section{Numerical Simulation of Silicon Bands}
\label{sec4}

\begin{figure}[tp]
    \centering
    \includegraphics[width=0.75\linewidth]{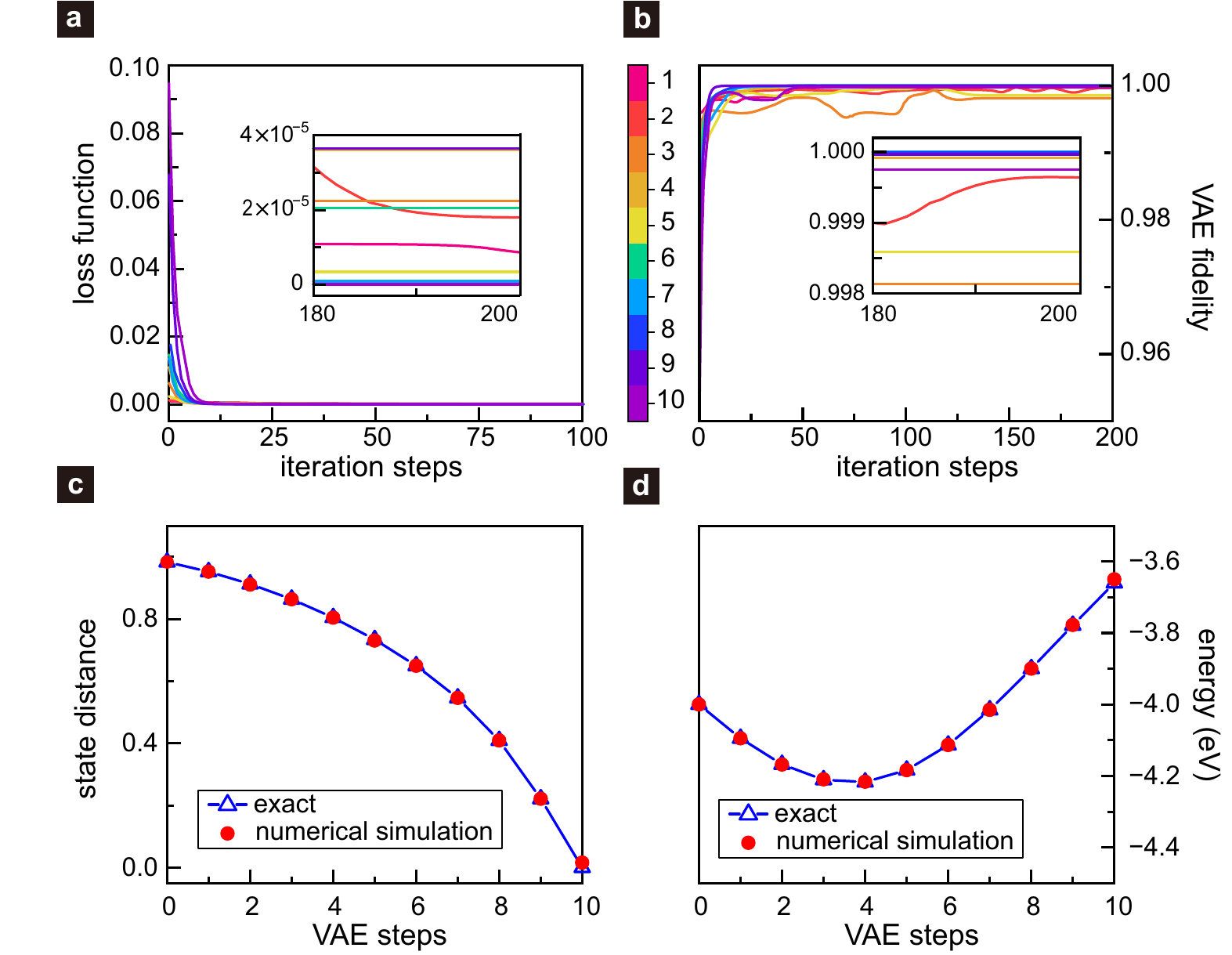}
    \caption{The numerical simulation of a 10-step VAE in the determination of the $W$-VB state $|\varphi^\mathrm{VB}_W\rangle$.
    At each $j(=1,\cdots, 10)$th VAE step, the evolution of the loss function $\mathcal F_\mathrm{num}({\boldsymbol{\theta}_j})$
    and the VAE fidelity $f_\mathrm{num}({\boldsymbol{\theta}_j})$ with the increase of the iteration step $s$ is shown in (a) and (b), respectively.
    The results of $\mathcal F_\mathrm{num}({\boldsymbol{\theta}_j})$ and $f_\mathrm{num}({\boldsymbol{\theta}_j})$
    for $180\mspace{-3mu}\le\mspace{-3mu} s\mspace{-3mu}\le\mspace{-3mu} 200$ are shown in the two insets.
    The results for different values of $j$ are displayed in different colors.
    With the increase of the VAE step, the distance $\mathcal D_j$ between the $j$th optimal VAE state $|\psi^\mathrm{V}_j\rangle^\mathrm{opt}_\mathrm{num}$
    and the target state $|\varphi^\mathrm{VB}_W\rangle$ is shown in (c), while the energy expectation
    $\mathcal E_j\mspace{-3mu}$ is shown in (d). 
    The blue solid lines with open triangles refer to the exact results from the matrix diagonalization
    while the red closed circles refer to the numerical results of the VAE. }
    \label{fig2}
\end{figure}

In this section, we present the numerical simulation of the VAE in computing
the electronic band structure of silicon~\cite{Hamaguchi2010book,Brust1964}.
This study focuses on the highest
valence band (VBI) and the lowest conduction band (CBI) around the Fermi
level, as they govern the key electronic properties of semiconductors.
As shown in Fig.~\ref{fig1}(c), the HE ansatz quantum circuit~\cite{KandalaNature17}
in our study includes 2.5 layers of the total 14 $Y$ gates and 6 C$Z$ gates.
The reference state is set to be the ground state, $|\psi_r\rangle\mspace{-3mu} =\mspace{-3mu}|0000\rangle$.

For each band, our VAE computation starts with the determination of
the Bloch states $\{|\varphi^\mathrm{VB}_W\rangle,|\varphi^\mathrm{CB}_W\rangle\}$
and their eigenenergies $\{\mathcal E^\mathrm{VB}_W,\mathcal E^\mathrm{CB}_W\}$
at the $W$-point $\vec{k}_W\mspace{-3mu} =\mspace{-3mu}(b/2,b,0)$ in the FBZ~\cite{Hamaguchi2010book}.
These two Bloch states (abbreviated to be the $W$-VBI and $W$-CBI states)
correspond to the 4th and 5th eigenstates of $H_W\mspace{-3mu} =\mspace{-3mu}H(\vec{k}_W)$ from Eq.~(\ref{eq2_3}).
We choose a diagonal initial Hamiltonian
$H_0\mspace{-3mu} =\mspace{-3mu}\sum_S \lambda_S |S\rangle\langle S|$.
The values $\lambda_S$ are slightly adjusted from the kinetic energies
$\{\lambda_{\vec{k}_W+\vec{G}}\}$ with the purpose of lifting any degeneracies present
in the free-electron basis.
For the $W$-VBI state, the initial eigenstate is $|\varphi_{n=4}(H_0)\rangle\mspace{-3mu} =\mspace{-3mu} |0011\rangle$
so that the 0th optimal VAE state is exactly given
by $|\psi_0\rangle_\mathrm{opt}\mspace{-3mu} =\mspace{-3mu}Y_3(\pi)Y_4(\pi)|\psi_r\rangle$.
 To be compatible with the 14-variable quantum circuit in Fig.~\ref{fig1}(c),
 an equivalent parameter set is given by
\be
{\boldsymbol{\theta}}^\mathrm{opt}_{j=0} &=& \{0,\pi/2,-\pi/2,-\pi/2,-\pi/2,\pi/2, \no \\
&& ~0,-\pi,\pi/2,0,0,-\pi,\pi/2,\pi/2 \}.
\label{eq4_1}
\ee
To adiabatically drag the state from $|\varphi_{4}(H_0)\rangle$ to $|\varphi_{4}(H_W)\rangle$,
a $10$-step VAE process is constructed
by the insertion of 9 breakpoints between $H_0$ and $H_W$, giving $H_j \mspace{-3mu} =\mspace{-3mu} (1-c_j) H_0 +c_j H_W$
with $c_{1\le j\le10} \mspace{-3mu} =\mspace{-3mu} j/10$.
At each $j$th VAE step, the condition of $|\langle \varphi_{4}(H_{j-1})|\varphi_{4}(H_j)\rangle|\mspace{-3mu} >\mspace{-3mu}97\%$
is expected to allow a reliable convergence of
$|\psi_j\rangle_\mathrm{opt}\mspace{-3mu} \approx\mspace{-3mu}|\varphi_{4}(H_j)\rangle$.
The evolution time involved in the loss function $\mathcal F(\calT; {\boldsymbol{\theta}}_j)$
is empirically selected around $\calT\approx 15~\h/\mathrm{eV}$.

\begin{figure}[t]
    \centering
    \includegraphics[width=0.75\linewidth]{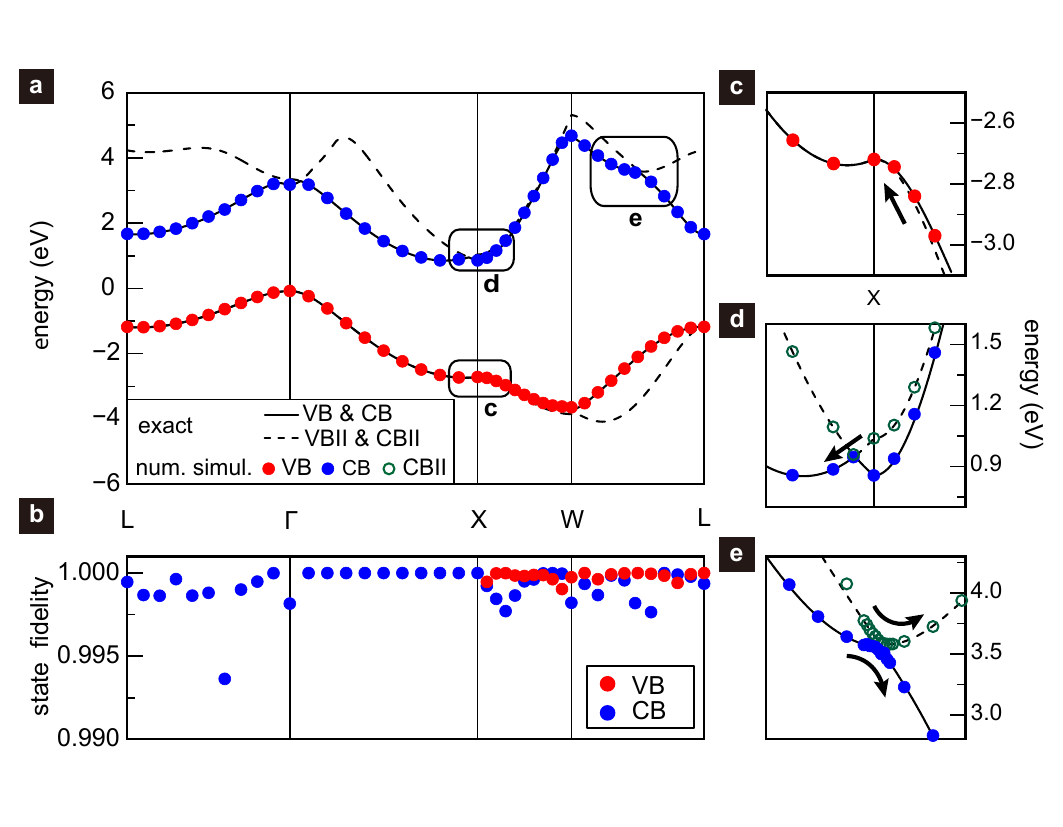}
    \caption{(a) The two electronic bands (VBI in red circles and CBI in blue circles)
    of silicon obtained by the numerical simulation of the VAE.
    As a comparison, their exact results from the matrix diagonalization are shown in solid lines,
    together with those of VBII and CBII are shown in dashed lines. To highlight special cases of the VAE,
    three regions of this band structure diagram are enlarged in (c), (d) and (e). The numerical results of CBII from the VAE
    in green open circles are also provided in (d) and (e). The fidelities between the optimal VAE states and the
    exact Bloch states are shown in (b). The red and blue  circles refer to the results of the VBI and CBI,
    respectively. The fidelities of the VB states between the $X$- and $L$-points are not calculated (see text
    for details).
      }
    \label{fig3}
\end{figure}

Figure~\ref{fig2} summarizes the numerical results of computing the $W$-VBI state.
At each $j$th VAE process step, the BFGS algorithm leads to a fast convergence of the loss function.
As shown in Fig.~\ref{fig2}(a), $\mathcal F(\calT; {\boldsymbol{\theta}}_{j=1,\cdots, 10})$
decreases dramatically within the first 10 iteration steps ($s\mspace{-3mu} \lesssim\mspace{-3mu} 10$).
To exhibit step-dependent results, we extend the iteration
up to 200 steps and $\mathcal F(\calT; {\boldsymbol{\theta}}_j)$ is stabilized in the order
of $10^{-5}$.
To further quantify the VAE performance, 
we calculate the evolution of the fidelity function~\cite{Nielsen2010book},
\be
f_{j; s} = \langle\varphi_{n}(H_j)| \psi_j({\boldsymbol{\theta}}^{(s)}_j)\rangle^2,
\label{eq4_2}
\ee
between the trial state $|\psi_j\rangle$ at the $s$th iteration step and the exact eigenstate $|\varphi_{n}(H_j)\rangle$
from the matrix diagonalization.
As shown in Fig.~\ref{fig2}(b), this fidelity function exhibits a fast increase toward its upper limit,
but small wiggling can be observed for large iteration steps ($s\mspace{-3mu}\gtrsim\mspace{-3mu}100$).
For simplicity, we assign the optimal parameters
with their values at the 200th step (${\boldsymbol{\theta}}^\mathrm{opt}_j\mspace{-3mu}=\mspace{-3mu}{\boldsymbol{\theta}}^{(s=200)}_j$)
and calculate the corresponding optimal state $|\psi_j\rangle_\mathrm{opt}$.
In our VAE process, the fidelity function of the optimal states consistently satisfies
$f_{j; \mathrm{opt}}\gtrsim 99.80\%$
and the final result toward $|\varphi^\mathrm{VB}_W\rangle$ is given by  $f_{j=10; \mathrm{opt}}\approx99.99\%$.
Empirical evaluation confirms that performing only 20 iterative steps per intermediate
point suffices to achieve adiabatic evolution to the target state at $W$-point.

To visualize the whole VAE process, we introduce the distance
$\mathcal D_{j}\mspace{-3mu}=\mspace{-3mu}[2-2\mathrm{Re}\langle \varphi^\mathrm{VB}_W|\psi_j\rangle_\mathrm{opt}]^{1/2}$
between the $j$th optimal state $|\psi_j\rangle_\mathrm{opt}$
and the final $W$-VBI state $|\varphi^\mathrm{VB}_W\rangle$.
As a comparison, the distances from the exact eigenstates
$\mathcal D^\pr_{j}\mspace{-3mu}=\mspace{-3mu}[2-2\mathrm{Re}\langle \varphi^\mathrm{VB}_W|\varphi_4(H_j)\rangle]^{1/2}$
are also calculated.
Figure~\ref{fig2}(c) shows that the state distance $\mathcal D_{j}$
systematically decreases from $\mathcal D_{0}\approx0.98$
to $\mathcal D_{10}\approx 0.01$. This evolution
agrees excellently with that of the exact value $\mathcal D^\pr_j$.
In addition, we inspect the evolution of the energy expectation
$\mathcal E_{j}\mspace{-3mu}=\mspace{-3mu} \langle \psi_j|H_j|\psi_j\rangle_\mathrm{opt}$,
which is plotted in Fig.~\ref{fig2}(d). An excellent agreement is also found between the energy expecation
$\mathcal E_j$ and the exact value from the matrix diagonalization.
The eigenenergy of the $W$-VBI state is determined at
$\mathcal E^\mathrm{VB}_{W}\mspace{-3mu}=\mspace{-3mu}-3.65 $ eV with a relative error of $0.15\%$.
The above VAE process is also applied to the $W$-CBI state (not shown),
which yields a similar convergence behavior.

Next we determine the band structure of silicon
by computing the dispersion relation $\mathcal E_{\vec{k}}$
for the $\vec{k}$-VBI and $\vec{k}$-CBI states~\cite{Hamaguchi2010book}.
As shown in Fig.~\ref{fig3}(a), we consider two special paths in the FBZ:
$W\mspace{-3mu}\rightarrow\mspace{-3mu} X \mspace{-3mu}
\rightarrow\mspace{-3mu} \Gamma \mspace{-3mu}\rightarrow\mspace{-3mu} L$ (path I)
and $W\mspace{-3mu}\rightarrow\mspace{-3mu} L$ (path II).
For both paths of the VBI states, the initial Hamiltonian and eigenstate are $H_0\mspace{-3mu}=\mspace{-3mu}H_W$
and $|\psi_0\rangle\mspace{-3mu}=\mspace{-3mu}|\psi_W\rangle_\mathrm{opt}\mspace{-3mu}\approx\mspace{-3mu}|\varphi^\mathrm{VB}_W\rangle$,
while the target Hamiltonian and eigenstate are $H\mspace{-3mu}=\mspace{-3mu}H_L$
and $|\varphi^\mathrm{VB}_L\rangle$  at the $L$ point.
To realize the VAE processes,
the total 29 breakpoints are inserted in path I while this number is 9 in path II.
The iteration steps ($s\mspace{-3mu}\leq\mspace{-3mu} 200$)
and the evolution time $\calT\approx 15~\h/\mathrm{eV}$ are consistent with those in
the treatment of the $W$-VBI and $W$-CBI states. 
The similar adiabatic evolution is designed for the CBI states along the two paths.
For the $j$th wavevector $\vec{k}_j$,
we obtain two optimal VAE states $|\psi_j\rangle_\mathrm{opt}$
and their energy expectations $\mathcal E_{\vec{k}_j}$ with the BFGS algorithm.
The results of $\mathcal E_{\vec{k}_j}$ are presented in Fig.~\ref{fig3}(a),
showing an excellent agreement ($\gtrsim\mspace{-2mu}99.50\%$ accuracy) with those from the exact diagonalization.
The outstanding performance of the VAE is also verified by
the fidelity function $f_{j; \mathrm{opt}}$ between the optimal
VAE state $|\psi_j\rangle_\mathrm{opt}$ and
the exact Bloch state $|\varphi_n(H_{\vec{k}_j})\rangle$.
As shown in Fig.~\ref{fig3}(b), the fidelity function satisfies
$f_{j; \mathrm{opt}}\mspace{-3mu}>\mspace{-3mu}99.70\%$
in most cases. This quantity is not calculated in the degeneracy region
(path $X\mspace{-3mu}\rightarrow\mspace{-3mu} L$ of the VBI states), which will be discussed below.

We previously used this anchor-point-based adiabatic dragging method
in a DAE study of 1D hydrogen chains~\cite{Zhan2021}.
Extending this approach to paths in three-dimensional (3D) momentum space,
however, demands greater care. Below, we examine key special
cases illustrated in Fig.~\ref{fig3}.
In the first scenario, the Bloch states of the VBI
and the second highest valence band (VBII) in the path
of $X\mspace{-3mu}\rightarrow\mspace{-3mu} L$
are degenerate in energy. As shown in Fig.~\ref{fig3}(c),
the VAE propagation along path I
converges to a linear combination of the VBI and VBII states.
Although the estimation of the eigenenergies is excellent,
the VAE optimal states evolves into a certain state within the
degenerate space.
The state fidelity $f_{j; \mathrm{opt}}$ becomes inconsistently
defined in this region and is not calculated in Fig.~\ref{fig3}(b).
In the second scenario, two nearby bands can accidentally cross
at certain wavevectors. For example, the CBI and the second
lowest conductance band (CBII) intersect around
the $X$-point. As shown in Fig.~\ref{fig3}(d),
the VAE propagation along path
I ($W\mspace{-3mu}\rightarrow\mspace{-3mu}\Gamma$)
of the CBI gives rise to the CBII states.
To overcome this problem, we design an additional
path ($W\mspace{-3mu}\rightarrow\mspace{-3mu}\Gamma$)
of the CBII to reach the correct CBI states.
In the third scenario, two nearby bands may not exactly intersect but
exhibit a very small band gap, e.g., the CBI and the CBII in the middle
region of path II. As shown in Fig.~\ref{fig3}(e), the evolution process
needs to proceed more slowly, so we added an extra 10 wave vectors in path II.
Then the VAE propagation can smoothly
pass this anti-crossing region and determine the CBI states.
Overall, the VAE paths must be carefully selected and certain modifications could be
necessary for an accurate band structure.

\section{Experimental Study of Silicon Bands }
\label{sec5}

In this section, we present our experimental study of silicon band
structures, building upon
the numerical simulations presented in Sec.~\ref{sec4}.
The experimental implementation is carried out on a 40-qubit superconducting
quantum processor, featuring a dual-row architecture as schematically
illustrated in Fig.~\ref{fig1}(b). The quantum control system
utilizes microwave $X$- and
$Y$-gates synthesized through an $I/Q$ modulator, which is driven by an
arbitrary waveform generator (AWG) board mixed with a microwave
source~\cite{ZhangIEEE2021}. Individual qubit frequencies are precisely tuned
via DC bias voltages applied to dedicated $Z$-control lines, which also enable
fast $Z$-rotations for single-qubit $Z$-gates.

To minimize wiring complexity, the $XY$ and $Z$ control signals for each
qubit are combined using a microwave diplexer mounted on the millikelvin-stage
flange, with the composite signal then routed to the quantum processor
through a single physical feedline. Tunable couplers provide adjustable
coupling strengths between intra-row neighboring qubits~\cite{YanPRAppl2018,BunpjQuantInf25},
while inter-row
couplers maintain fixed coupling strengths. Frequency detuning is maximized
between each inter-row qubit pair to suppress
unwanted idle-state interactions~\cite{WangPRApp19}.

Qubit state measurement is performed using a 3-$\mu$s low-power readout
pulse that enters through input port $R_\mathrm{in}$ and exits via
output port $R_\mathrm{out}$ after interacting with the qubits~\cite{ZhangIEEE2021,BunpjQuantInf25}.
The output signal undergoes two-stage cryogenic amplification before
being processed by a room-temperature data acquisition (DAQ) system~\cite{ZhangIEEE2021}.
This comprehensive experimental setup provides the necessary hardware
foundation for implementing the multi-qubit parameterized circuits
required for the VAE-based determination of silicon band structures,
combining precise quantum control with high-fidelity measurement
capabilities.

As illustrated in Fig.~\ref{fig1}(b), a subnetwork of four
qubits---labeled as qubit 27 ($Q_1$), 28 ($Q_2$), 8 ($Q_3$),
and 9 ($Q_4$)---is
selected to emulate the 4-qubit Hilbert space required for simulating
the electronic structure of silicon. These qubits exhibit relaxation
times of approximately $\sim100~\mu$s and pure dephasing times of
about $\sim 40~\mu$s (see Table 1 in Appendix), providing a coherence window sufficient for
multi-gate quantum circuits.
In alignment with the numerical simulations, the reference state
for the VAE is chosen as the
ground state $|\psi_r\rangle\mspace{-3mu}=\mspace{-3mu} |0000\rangle$.
At the beginning of each experimental run, the qubit system is fully
relaxed to this ground state. The ansatz circuit $U^\mathrm{V}({\boldsymbol{\theta}})$,
comprising 14 single-qubit $Y$ gates and 6 two-qubit $CZ$ gates
as depicted in Fig.~\ref{fig1}(c), is experimentally realized using a
multi-channel AWG. For clarity,
the VAE step index $j$ is omitted throughout this section unless explicitly required.

\begin{figure}[tp]
    \centering
    \includegraphics[width=0.75\linewidth]{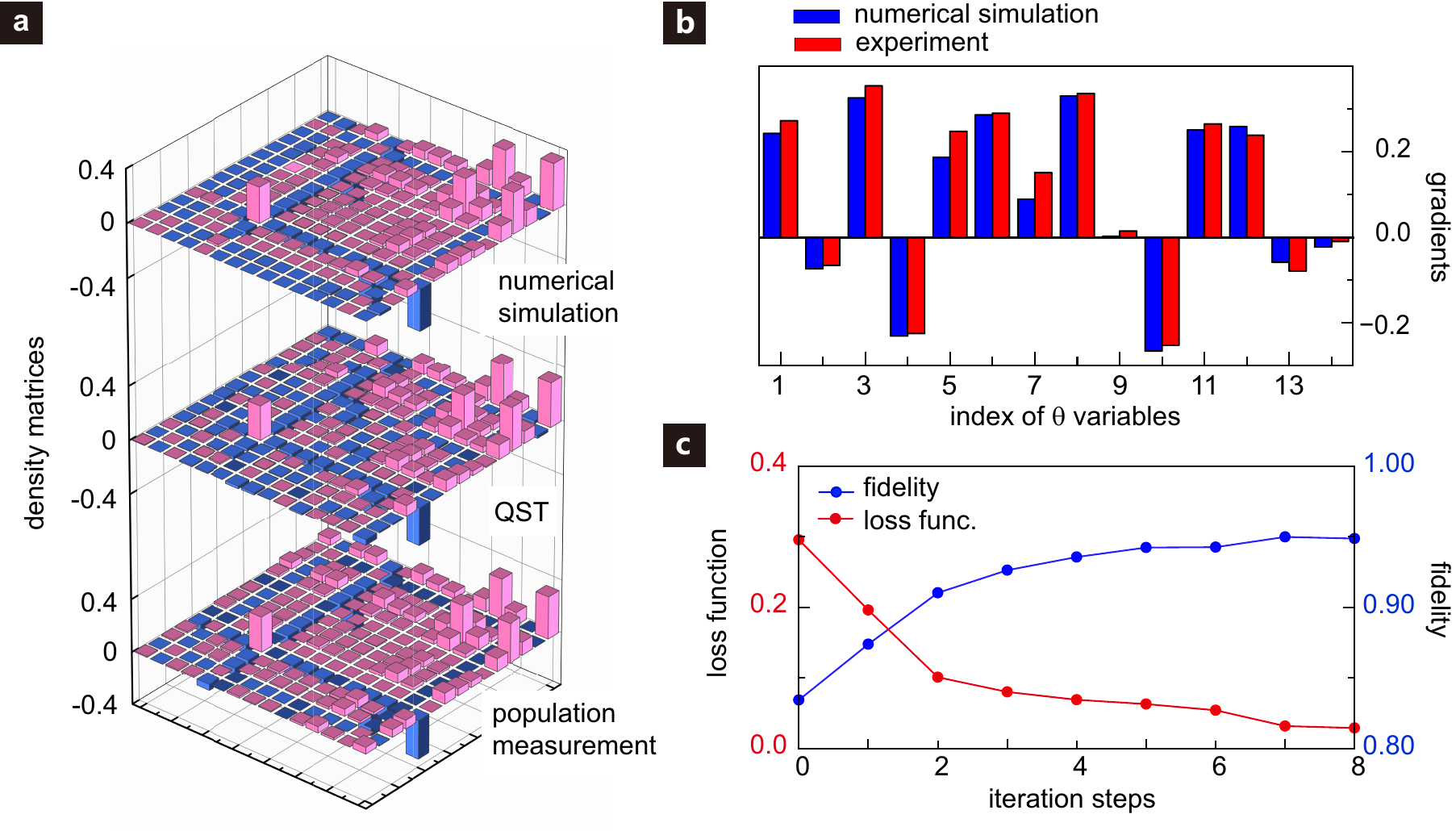}
    \caption{Experimental verification of the VAE protocol. (a) The $15\times 15$
    density matrices of the optimized VAE state $|\psi^\mathrm{V}\rangle^\mathrm{opt}$
    for the $W$-VB state, comparing numerical simulation of
    $|\psi^\mathrm{V}\rangle^\mathrm{opt}_\mathrm{num}$ (top), experimental quantum
    state tomography of $|\psi^\mathrm{V}\rangle^\mathrm{opt}_\mathrm{exp}$ (middle),
    and single-shot population measurement of
    $|\psi^\mathrm{V}\rangle^\mathrm{opt}_\mathrm{exp}$ (bottom).
    The experimental state is prepared using numerically optimized parameters
    ${\boldsymbol{\theta}}^\mathrm{opt}_W$ applied to the ansatz circuit.
    Matrix elements are color-mapped with positive (pink) and negative (dark blue) values.
    (b) Partial derivatives $\partial \mathcal F/\partial \theta_{l=1, \cdots, 14}$
    of the loss function evaluated at the first iteration ($s=1$) of the final
    VAE step ($j=10$) for the determination of $|\varphi^\mathrm{VB}_W\rangle$.
    Numerical (blue) and experimental (red) results are shown for each of the
    14 parameters.
    (c) Experimental implementation of the VAE with gradients and loss extracted
    from single-shot quantum measurements. The optimization trajectory
    from $H_9$ to $H_W$ is shown, depicting the loss (red) and the fidelity
    with the exact $W$-point eigenstate (blue) as functions of iteration step. }
    \label{fig4}
\end{figure}

Gate-level implementation details are as follows: each single-qubit $Y$-rotation
gate $Y_{i}(\theta_l)$ is executed using a microwave pulse with a cosine-shaped
envelope, where the rotation angle $\theta_{l}$ is controlled via the pulse
amplitude. Each two-qubit controlled-$Z$ gate $\mathrm{C}Z_{i_1,i_2}$
is implemented using synchronized square pulses applied to both the
control ($i_1$) and target ($i_2$) qubits~\cite{BunpjQuantInf25}.
The measured gate fidelities are approximately 99.9\% for $Y$ gates and 99.0\% for
C$Z$ gates (see Table 1 in Appendix). Despite the relatively large number of gates, the total
duration of the experimental circuit is approximately 600 ns, which
remains significantly shorter than the qubit coherence times.
Following the preparation of a trial state
$|\psi\rangle_\mathrm{exp}\mspace{-3mu}=\mspace{-3mu}U^\mathrm{V}({\boldsymbol{\theta}})|\psi_r\rangle$,
population measurements are performed via dispersive readout.
Here, all experimentally generated states are labeled with
the superscript `exp'. Each measurement cycle is repeated 50,00 times
to accurately estimate the population
$P_S\mspace{-3mu}=\mspace{-3mu}\langle S|\psi^V\rangle^2_\mathrm{exp}$
for each computational basis state $|S\rangle\mspace{-3mu}=\mspace{-3mu}|s_1s_2s_3s_4\rangle$.
The dispersive readout introduces inherent assignment errors,
ranging from $1\%\sim5\%$ for the ground state and $5\%\sim10\%$
for excited states (see Table 1 in Appendix). These errors are substantially reduced
through the application of a readout correction matrix,
as detailed in Appendix.

To fully characterize the experimentally prepared trial state $|\psi\rangle_\mathrm{exp}$,
quantum state tomography (QST) is generally employed~\cite{Nielsen2010book,BunpjQuantInf25}.
The density matrix $\rho_\mathrm{exp}\mspace{-3mu}=\mspace{-3mu}|\psi\rangle\langle\psi|_\mathrm{exp}$
is linearly decomposed over the complete set of Pauli operators as
$\rho_\mathrm{exp}\mspace{-3mu}=\mspace{-3mu}\sum_\alpha \varrho_\alpha \mathcal{O}_\alpha$,
where the expansion coefficients are given by
$\varrho_\alpha\mspace{-3mu}=\mspace{-3mu}\langle\psi|\mathcal O_\alpha|\psi\rangle_\mathrm{exp}$.
Each coefficient $\varrho_\alpha$ is determined by measuring the population of an
alternatively prepared state
$|\psi\rangle_{\alpha} \mspace{-3mu}=\mspace{-3mu} U_\alpha |\psi\rangle_\mathrm{exp}$,
generated by applying a specific pre-rotation gate $U_\alpha$ prior
to the standard computational basis measurement. A full reconstruction of the density
matrix requires approximately 80 distinct population measurements (see Appendix).

As a demonstrative example, we experimentally prepare and measure the
optimal VAE state $|\psi\rangle_{\mathrm{opt}}^{\mathrm{exp}}\mspace{-3mu}=\mspace{-3mu}U^\mathrm{V}(\boldsymbol{\theta}^\mathrm{opt}_W)|\psi_r\rangle$.
Figure~\ref{fig4}(a) compares the experimentally reconstructed density
matrix $\rho_\mathrm{exp}$, obtained via quantum state tomography (QST),
with the theoretically predicted density matrix $\rho$ computed
from numerical simulation.
Note that, owing to the real-valued Hamiltonian, only the real components
of the density matrix has been measured, and the imaginary components
are artificially omitted.
Visually, the two matrices show strong similarity in their major components.
The QST fidelity, defined as
$f\mspace{-3mu}=\mspace{-3mu}\mathrm{tr}\{\rho_\mathrm{exp}\rho\}$,
is calculated to be 92.7\%.

Given that the effective silicon Hamiltonian (Eq.~\ref{eq2_2}) is
real-valued, the expansion coefficients $c_S$ of its eigenstates
in the computational basis are also real. This property enables
a simplified state determination method: we directly measure the
populations $P_S\mspace{-3mu}=\mspace{-3mu}c^2_S$ in a single
experimental shot, and reconstruct the state as
$|\psi\rangle_\mathrm{exp}\mspace{-3mu}=\mspace{-3mu}\sum_S s_S \sqrt{P_S}|S\rangle$,
where the sign $s_S\mspace{-3mu}=\mspace{-3mu}1$ or $-1$
for each basis state is assigned by comparing with the relative
phase information from numerical simulation. As also
shown in Fig.~\ref{fig4}(a), the density matrix obtained via
this simplified approach agrees well with the theoretical prediction,
yielding an even higher state fidelity of $f\mspace{-3mu}=\mspace{-3mu}97.6\%$.
This improvement is attributed to the avoidance of error accumulation
inherent in the extensive measurement sequences required for full
QST. Consequently, all experimental
states reported in this work are characterized using this efficient
single-shot measurement method.

A pivotal technical component of VAE is the gradient-based optimization method
employed to locate the global minimum of the phase-based loss function
$\mathcal F(\calT; {\boldsymbol{\theta}})$~\cite{Nocedal2016book}.
To assess the experimental feasibility of the VAE protocol, we
directly measure the full gradient
$\nabla\mathcal F(\calT; {\boldsymbol{\theta}})$ on the quantum processor.
As a representative demonstration, we consider the specific parameter set
${\boldsymbol{\theta}}\mspace{-3mu}=\mspace{-3mu}{\boldsymbol{\theta}}^{(s=0)}_{j=10}$,
corresponding to the initial iteration ($s=0$) of the final VAE step ($j=10$)
in the preparation of the $W$-point valence band (VBI)
state (refer to Fig.~\ref{fig2} for the context).
The complete gradient vector comprising all 14
elements $\partial \mathcal F/\partial \theta_{l=1,\cdots, 14}$ is shown in
Fig.~\ref{fig4}(b), comparing results from numerical simulation and experimental
measurement. A good qualitative agreement is observed between the theoretical
and experimentally measured gradients.
Then we employ single-shot measurements to characterize the quantum state
at each iteration, compute the parameter gradients, and update
the variational parameters accordingly. This process steers
the state toward the target eigenstate at $W$-point. As shown in
Fig.~\ref{fig4}(c), starting from the parameters of the 9th intermediate
point for $H_9$, the fidelity between the experimentally prepared state and the
ideal eigenstate of $H_W$ rises from approximately 85\% to 95\% over eight
gradient-based updates. The corresponding loss function decays
monotonically, as depicted in Fig.~\ref{fig4}(c). The final loss function
at the 8th step is several orders larger than the converged simulation
value, primarily originating from device-level errors, including readout
errors and gate errors. Figure~\ref{fig4}(c) provides direct experimental
verification of gradient-based quantum optimization.

\begin{figure}[tp]
    \centering
    \includegraphics[width=0.75\linewidth]{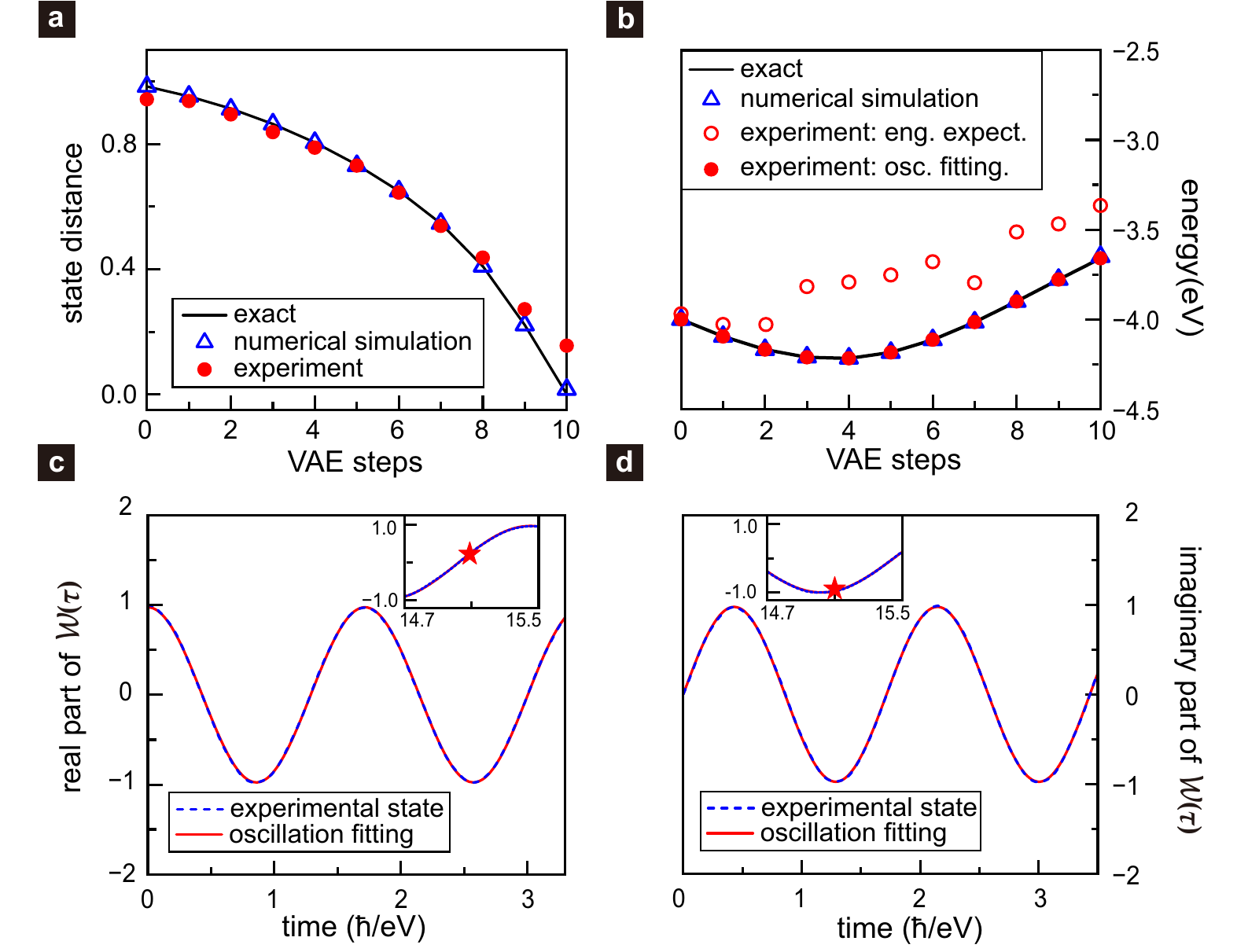}
    \caption{Experimental validation of the VAE for determining the $W$-VB state (refer to Fig.~\ref{fig2}).
    (a) State distance $\mathcal D_j$ between the optimal VAE state
    at step $j(=\mspace{-3mu}0,\cdots, 10)$ and the target $W$-VB eigenstate.
    The black solid line denotes results from exact eigenstates, blue triangles
    and red circles correspond to numerical and experimental estimates of
    $|\psi^\mathrm{V}_j\rangle^\mathrm{opt}$, respectively.
    (b) Evolution of the $j$th eigenenergy. The black solid line shows exact
    results; blue triangles and red open circles represent energy expectations
    from numerical state $|\psi^\mathrm{V}_j\rangle^\mathrm{opt}_\mathrm{num}$
    and experimental optimal states
    $|\psi^\mathrm{V}_j\rangle^\mathrm{opt}_\mathrm{exp}$. Red closed circles
    indicate results from single-frequency fitting of the time-evolution
    expectation $\mathcal W_{j; \mathrm{exp}}(\tau)$.
    (c, d) Real and imaginary parts of $\mathcal W_{j=10; \mathrm{exp}}(\tau)$ (blue
    dashed) and the corresponding fitted curve
    $\mathcal W_{j=10; \mathrm{fit}}(\tau)$ (red solid). Red stars mark
    $\mathcal W_{j=10}(\tau_\mathrm{V})$ used in optimizing
    ${\boldsymbol{\theta}}_{j=10}$.}
    \label{fig5}
\end{figure}

We now experimentaly validate the 10-step VAE protocol
for preparing the valence band ($W$-VBI) state at the $W$-point
($|\varphi^\mathrm{VB}_W\rangle$).
To utilize limited quantum resources and mitigate error
accumulation, we directly generate and
measure all 11 optimized VAE states
$|\psi_j\rangle_\mathrm{opt}^\mathrm{exp}$ (with parameters
${\boldsymbol{\theta}}^\mathrm{opt}_{j=0, \cdots, 10}$)
along the adiabatic path. These experimentally
prepared states consistently achieve high fidelities of
approximately 98\% relative to their numerical
counterparts (detailed values not shown here).
To quantitatively evaluate the VAE process, we examine
the state evolution by calculating the distance
$\mathcal D_j^{\mathrm{exp}}$ between each experimentally
prepared state $|\psi_j\rangle_\mathrm{opt}^\mathrm{exp}$
and the target $W$-VBI state $|\varphi^\mathrm{VB}_W\rangle$~\cite{Nielsen2010book}.
As shown in Fig.~\ref{fig5}(a), the progression of
$\mathcal D_{j}^{\mathrm{exp}}$ closely follows the theoretical simulation,
demonstrating the effectiveness of the VAE approach.

We further analyze the energy expectations
$\mathcal E_{j}^{\mathrm{exp}}\mspace{-3mu}=\mspace{-3mu}\langle\psi_j|H_j|\psi_j\rangle_\mathrm{opt}^\mathrm{exp}$
throughout the VAE process. While these values provide a qualitatively
correct description of the eigenenergies $\mathcal E_{j}$ [Fig.~\ref{fig5}(b)],
systematic deviations of approximately 0.5 eV are present.
These discrepancies arise from minor contaminations of
high-energy eigenstates in the experimentally prepared states.
To mitigate these errors, we employ a phase-based characterization
method. We compute the expectation value of the time evolution
operator
$\mathcal W_\mathrm{exp}(\tau)
\mspace{-3mu}=\mspace{-3mu}\langle\psi_j|\mathcal U_j(\tau)|\psi_j\rangle_\mathrm{opt}^\mathrm{exp}$
over an extended time domain $\tau$. A single-frequency fitting
function $\mathcal W_\mathrm{fit}(\tau)\mspace{-3mu}=\mspace{-3mu}A\exp(-i\omega_j \tau)$
effectively captures the dominant eigenstate contribution,
as demonstrated in Fig.~\ref{fig5}(c) and \ref{fig5}(d) for the final step ($j=10$).
The excellent agreement between $\mathcal W_\mathrm{fit}(\tau)$
and $\mathcal W_\mathrm{exp}(\tau)$ yields fitting parameters
$A\mspace{-3mu}=\mspace{-3mu}0.976$ and
$\omega_{j=10}\mspace{-3mu}=\mspace{-3mu}-3.66~\mathrm{eV}/\hbar$.
The coefficient $A$ represents the fidelity between the experimental
state and the target $W$-VBI state, while the frequency $\omega_{j=10}$
provides an improved estimation of the eigenenergy
$\mathcal E_{j=10}^{\mathrm{fit}}\mspace{-3mu}=\mspace{-3mu}\hbar\omega_j\mspace{-3mu}=\mspace{-3mu}-3.66$ eV.
The evolution of the fitted eigenenergies
$\mathcal{E}_j^{\mathrm{fit}}$ ($j = 0, 1, \dots, 10$) across all
VAE iterations [Fig.~\ref{fig5}(b)] demonstrates markedly improved
agreement with the theoretical values, reducing the absolute
deviation to less than 0.01 eV. This quantitatively
validates the enhanced accuracy of our phase-based
characterization approach.

\begin{figure*}[t]
    \centering
    \includegraphics[width=0.55\linewidth]{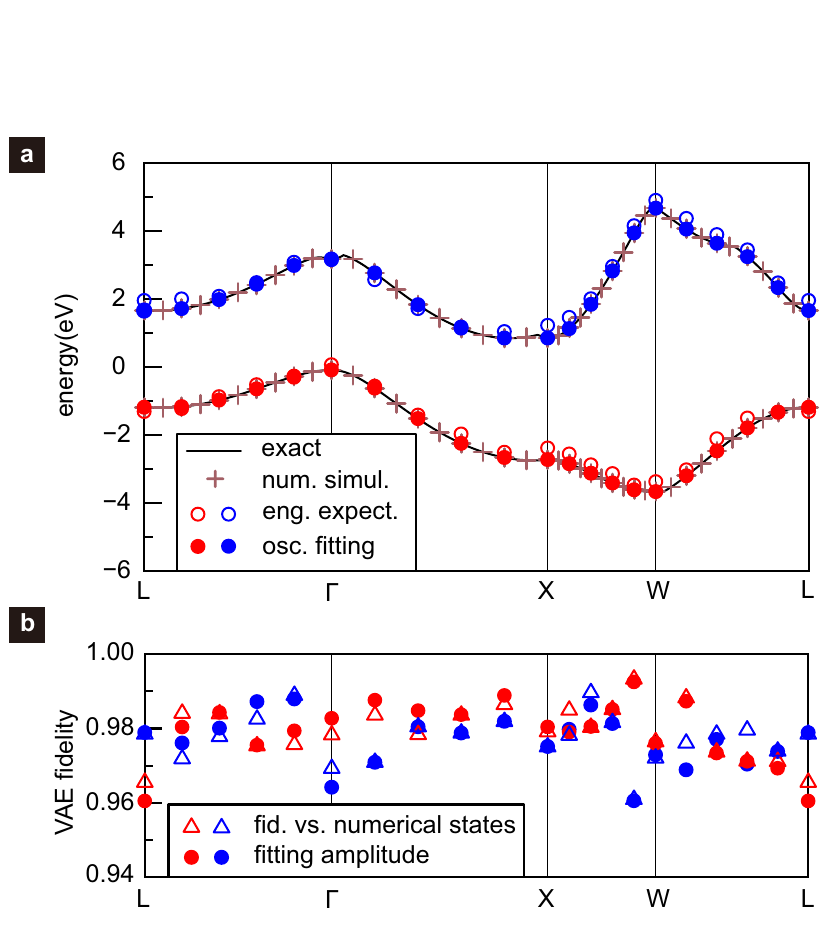}
    \caption{Experimental determination of the silicon band structure using VAE.
    (a) Valence and conduction bands along selected symmetry paths.
    Black solid curves: exact results; brown crosses: VAE predictions
    from numerical simulation. Open and filled circles denote experimental
    estimates from direct energy expectation and single-frequency fitting,
    respectively.
    (b) Fidelities of experimentally prepared VAE states. Open triangles:
    fidelity relative to numerical simulation; filled circles: fidelity
    extracted from the fitted amplitude $A_{\vec{k}}$. In both panels,
    red and blue symbols correspond to valence and conduction bands, respectively. }
    \label{fig6}
\end{figure*}

Having validated the VAE methodology, we apply the complete
experimental protocol---encompassing state generation, measurement, and
advanced data analysis---to determine the full electronic band structure
of silicon. A set of 31 wavevectors $\vec{k}$ are selected along
high-symmetry paths (I and II) in the first Brillouin zone,
as illustrated in Fig.~\ref{fig3}. For each wavevector,
the VAE process is implemented experimentally with optimized sets
${\boldsymbol{\theta}}^\mathrm{opt}_{\vec{k}}$.
The corresponding optimal VAE states are denoted as valence band
states $|\psi_{\vec{k}}^\mathrm{VB}\rangle_\mathrm{opt}^\mathrm{exp}$
and conduction band states
$|\psi_{\vec{k}}^\mathrm{CB}\rangle_\mathrm{opt}^\mathrm{exp}$.
These experimental states exhibit consistently high fidelities
exceeding 96\% compared to their numerical counterparts
across all wavevectors, as shown in Fig.~\ref{fig6}(b),
demonstrating the reliability of both the state generation
protocol and the simplified single-shot measurement approach.

The energy expectations
$\mathcal E_{\vec{k}}^{\mathrm{exp}}\mspace{-3mu}=\mspace{-3mu}
\langle\psi_{\vec{k}}|H(\vec{k})|\psi_{\vec{k}}\rangle_\mathrm{opt}^\mathrm{exp}$
computed for all sampled wavevectors provide a qualitatively
correct description of the band structure, as presented in
Fig.~\ref{fig6}(a). However, these direct measurements exhibit systematic
deviations of up to 0.5 eV due to spectral contamination
from non-dominant eigenstates. To achieve higher accuracy,
we implement a spectral
fitting procedure using the single-frequency oscillation model
$\mathcal W_{\vec{k}}^{\mathrm{fit}}(\tau)\mspace{-3mu}=\mspace{-3mu}A_{\vec{k}}\exp(-i\omega_{\vec{k}}\mspace{1.25mu}\tau)$.
The refined energy estimates
$\mathcal{E}_{\vec{k}}^{\mathrm{fit}} = \hbar\omega_{\vec{k}}$
exhibit excellent agreement with the exact eigenenergies
in Fig.~\ref{fig6}(a), also reducing the absolute deviation
to less than 0.01 eV; this improvement arises because the
fitting procedure effectively suppresses spurious
contributions from non-dominant eigenstates.
As a metric of state fidelity, the fitted amplitude
$A_{\vec{k}}$ in Fig.~\ref{fig6}(b) exceeds 96\% and,
importantly, quantifies the overlap between the experimental
and target eigenstates, without requiring prior knowledge of the exact eigenstate.
This property renders the metric robust for analyzing degenerate regions
like the $X\mspace{-3mu}\rightarrow\mspace{-3mu} L$ path, where the
target is a linear combination of degenerate
eigenstates.
These comprehensive results demonstrate
that the VAE protocol combined with spectral analysis enables
reliable band structure determination on quantum processors,
achieving both high state fidelity and accurate energy resolution.

\section{Discussion and Conclusion}
\label{sec6}

In this work, we have designed and implemented a VAE
for determining eigenstates of
multi-qubit Hamiltonians. The VAE integrates the conceptual
framework of adiabatic evolution with the flexibility of
variational quantum circuits, offering a practical approach
for eigenstate preparation on NISQ
devices. By combining a hardware-efficient ansatz with a
phase-based loss function, the VAE overcomes key limitations
of conventional analog and digitized adiabatic methods
while maintaining high fidelity in eigenstate preparation.

We have implemented the VAE to compute the electronic
band structure of silicon, with particular emphasis on
the highest valence band (VBI) and the lowest
conduction band (CBI). Under a single-electron framework,
the effective Hamiltonian is encoded into a 4-qubit system.
Our computational approach involves selecting high-symmetry
anchor points in the Brillouin zone---such as the $W$-point---and
propagating variationally optimized states along symmetric
paths including $W$--$X$--$\Gamma$--$L$ and $W$--$L$ within
the first Brillouin zone. A hardware-efficient ansatz
circuit comprising 2.5 layers and 14 parameters has been
utilized, initialized from the 4-qubit ground state.

Numerical simulations demonstrate excellent performance,
achieving state fidelities mostly above 99.7\% and energy
accuracies exceeding 99.5\%. Experimental implementation
on a superconducting quantum processor yields state
fidelities above 96\% with single-shot
readout and around 92\% via quantum state tomography.
Even with experimentally estimated gradients, the optimized state
can reach a high fidelity of approximately 95\% to the target eigenstate.
The VAE protocol consistently
produces eigenenergy estimates within approximately
0.5 eV of theoretical values. Further refinement
using single-frequency oscillation fitting reduces the
deviation to less than 0.01 eV, underscoring
the robustness of the method even under experimental
imperfections.

Our VAE integrates adiabatic and variational approaches
to overcome their respective limitations. While adiabatic
methods (AAE/DAE) provide deterministic eigenstate
preparation, they require deep circuits susceptible
to hardware noise. Variational approaches like VQE
employ parameterized circuits but struggle with
excited-state convergence in complex energy
landscapes~\cite{TillyPhysRep22,Xie2022}.
For instance, variance-VQE
faces challenges in identifying suitable initial
parameters that guarantee convergence to target
excited states. The VAE addresses these
issues by incorporating
adiabatic guidance to facilitate variational
excited-state search. This synergy creates a
scalable, noise-resilient strategy for band
structure calculations, representing a general
paradigm for enhancing quantum algorithms. While
previous hybrid adiabatic-variational approaches have focused predominantly on ground-state preparation~\cite{GarciaSaezLatorre18,Matsuura2020,Matsuura2021,Harwood2022,Schiffer2022},
our VAE framework is extended via a phase-based
loss function that enables unbiased treatment
of excited states. This capability is experimentally demonstrated by
reconstructing the silicon valence and conduction band structure.

In this framework, the adiabatic pathway
serves as a foundational component rather
than an auxiliary tool, ensuring reproducible
excited-state preparation. Our implementation
specifically addresses experimental constraints
through customized Hamiltonian construction,
circuit design, and path planning for superconducting
hardware. We expect that the adiabatic trajectory
guidance can alleviate the barren plateau
problem~\cite{McCleanNatCom18} common in variational algorithms,
reducing local minima convergence risks
and enhancing optimization stability.

While the VAE framework shows considerable promise,
practical challenges must be addressed to
advance its applicability. Gradient-based iterations
in standard VQE require repeated estimation of the
Hamiltonian expectation $\langle H \rangle$, usually
obtained by measuring each Pauli term in the operator
expansion. A key limitation of this $\langle H \rangle$
strategy is its poor transferability to excited-state
calculations: the corresponding cost functions
are generally not equivalent to minimizing a
single expectation. Here, we instead employ a loss
function defined directly by single-shot measurements
of the final quantum state. For the real-value Hamiltonian
used in this experiment, sign information of the probability
amplitudes has been obtained from classical simulation,
improving data efficiency. For general complex Hamiltonians,
a universal state-based measurement scheme remains
necessary---e.g., simplified QST~\cite{TanDianPRL2024},
or protocols using N auxiliary qubits (total 2N) with a
controlled-gate operation to read out the
complete state in parallel~\cite{YouZhouPRAppl2025}.
For the dense Hamiltonian considered, the
Pauli-term count is on the same order as the resources
required for full state tomography. Hence, efficient
final-state measurement techniques are of immediate
value to our scheme and could also provide good estimation of $\langle H \rangle$
in general VQE settings.

In the future, we also expect to directly evaluate the
loss-function term $\langle e^{-i H \mathcal{T}/\hbar} \rangle$
within the experimental quantum circuit. This can be implemented
via a controlled-$e^{-i H \mathcal{T}}$ evolution: an ancilla qubit,
prepared with a Hadamard gate, controls the application of the
time-evolution operator on the main register~\cite{Schiffer2022}. Measuring this ancilla
then extracts the relevant expectation of evolution---and thus the loss---through
the ancilla-qubit readout, bypassing the corresponding classical
post-processing.

Scaling the current scheme to larger systems also
encounters challenges of growing circuit depth and limited
expressibility in the expanded Hilbert space---which must
be systematically evaluated and addressed via concrete
numerical and experimental benchmarks.
Furthermore, extending the method beyond single-electron
models to include many-body electron-electron
interactions will be essential for simulating
realistic quantum materials. With continued progress
in quantum hardware and algorithmic design, the
VAE offers a viable and extensible pathway
toward high-accuracy simulation of complicated
quantum systems.

\section*{Acknowledgements}
The authors from Zhejiang University (Y.Y. and colleagues) thank the
support from the National Natural Science Foundation of China (Grant No. 12074336),
the National Key Research and Development Program of
China (Grant No. 2025YFH0102104, and No. 2022YFA1403202). Y.Y. also acknowledges the funding
support from Tencent Corporation.

\section*{Data Availability}
The data that support the findings of this article are
openly available~\cite{DataSet}.

\section*{APPENDIX: METHODS}
\subsection*{1. Readout Correction}

We implement the VAE protocol on a superconducting Xmon
processor to determine silicon's band structure. Qubit parameters
are listed in Table 1. Using dispersive readout, we infer qubit
states from resonator frequency shifts, a process inherently
susceptible to assignment errors requiring systematic correction.

\begin{table*}[h]
\centering
\tabcolsep=0.3cm
\begin{tabular}{|c|c|c|c|c|c|c|c|}
\hline          & $f_{10}$ (GHz) & $T_1 $($\mu$s)& $T_2^*$($\mu$s)  & $F_g$ & $F_e$ & $F_{sg}$ & $F_{tg}$\\
\hline  $Q_1$   & $ 4.83 $ &$60.79$ & $24.46$ & $99.06\%$ & $94.79\%$ & $99.95\%$   & $99.4\%$(with Q2)     \\
\hline  $Q_2$   & $ 4.94 $ &$96.53$ & $55.16$ & $98.96\%$ & $94.25\%$& $99.95\%$  & $99.3\%$(with Q3)       \\
\hline  $Q_3$   & $ 4.38 $ &$98.62$ & $46.96$ & $96.75\%$ & $90.74\%$ & $99.97\%$  & $99.4\%$(with Q4)      \\
\hline  $Q_4$   & $ 4.51 $ &$66.7$ & $57.11$ & $95.53\%$ & $91.69\%$ & $99.95\%$   & $ $      \\
\hline
\end{tabular}
\caption{Key parameters of the four superconducting qubits used in this work,
including frequencies ($f_{10}$), relaxation times ($T_1$ and spin-echo $T_2^* $),
readout fidelities for the ground ($F_g$) and excited ($F_e$) states,
and fidelities of single- ($F_{sg}$) and two-qubit ($F_{tg}$) gates.
}
\end{table*}

The readout correction procedure is illustrated through a single-qubit example.
When a qubit is prepared in state $|0\rangle$, the measurement yields $|0\rangle$
with probability $P(0|0)$ and $|1\rangle$ with
probability $P(1|0)=1-P(0|0)$. Similarly, preparing $|1\rangle$ gives
outcomes $|0\rangle$ and $|1\rangle$ with probabilities $P(0|1)$ and $P(1|1)=1-P(0|1)$,
respectively. For a general
state $|\psi\rangle = c_0 |0\rangle + c_1 |1\rangle$, the true
populations $P_0=|c_0|^2$ and $P_1=|c_1|^2$ relate to the
measured populations $P_0^\pr$ and $P_1^\pr$ via:
\be
P^\pr_0 &=& P(0|0) P_0 + P(0|1) P_1,  \no \\
P^\pr_1 &=& P(1|0) P_0 + P(1|1) P_1.
\label{eq_s10}
\ee
In vector form, $\mathbf{P}^\pr = R \mathbf{P}$, where
$\mathbf{P} = (P_0, P_1)^T$, $\mathbf{P}^\pr = (P_0^\pr, P_1^\pr)^T$,
and the readout matrix is:
\be
	R=\left(\begin{array}{ll}
		P(0|0) & P(0|1) \\
		P(1|0) & P(1|1)
	\end{array}\right).
\label{eq_s12}
\ee
Assuming $R$ is constant, the true population is recovered
as $\mathbf{P} = R^{-1}\mathbf{P}^\pr $.

This measurement correction method can be straightforwardly extended to the
multi-qubit cases. For an $N$-qubit device, one simple approach
is to assume that each qubit works independently so that the total readout
matrix is written as $R=R_{1} \otimes R_{2} \otimes \cdots \otimes R_{N}$,
where $R_i$ is the readout matrix for each $i$th qubit.
An alternative is to generate each basis state
$|S\rangle = |s_1\cdots s_N\rangle$ with $|s_{i=1, \cdots, N}\rangle \in \{|0\rangle_i, |1\rangle_i\}$,
and perform the population measurement of all the basis states $\{|S^\pr\rangle\}$.
The resulted conditional probabilities of $\{P(S^\pr|S)\}$ are used to
construct the $2^N\times 2^N$ readout matrix $R$ with
$R_{S^\pr, S} = P(S^\pr|S)$. Then we can estimate the real
population vector $\mathbf{P}$ from the measured population
vector $\mathbf{P}^\pr$ and $R_{S^\pr, S}$.

To evaluate the method, we apply it to the $W$-point valence band
state obtained via a 10-step VAE protocol on four qubits.
At each step $j$, the optimal VAE state
$|\psi^\mathrm{V}_{j}\rangle^\mathrm{opt}$ is prepared experimentally,
and the fidelity with the numerical state is computed as:
\be
\mathcal F_j = (\sum_{S} \sqrt{P^{\mathrm{exp}}_{j; S}}\sqrt{ P^{\mathrm{num}}_{j; S}})^2,
\label{eq_s16}
\ee
where $P^{\mathrm{exp}}_{j; S}$ and $P^{\mathrm{num}}_{j; S}$ are
the numerical and experimental population  along each basis state $|S\rangle$.
After readout correction, the fidelity improves consistently
from $0.7<\mathcal F_j< 0.85$ to be around $0.97$ (not shown), confirming the
effectiveness of the readout correction.


\subsection*{2. Quantum State Tomography}

To fully characterize an experimentally prepared quantum state,
one must construct its density matrix by measuring both population
and coherence information. While dispersive readout provides only
population data, quantum state tomography (QST) enables complete
state reconstruction through Pauli basis expansion.

For a single qubit, the density matrix expands as:
\be
\rho = \varrho_I I + \varrho_X X +  \varrho_Y Y + \varrho_Z Z,
\label{eq_s19}
\ee
where coefficients
$\varrho_{\sigma}=\mathrm{Tr}\{\rho\sigma\}=\langle\psi|\sigma|\psi\rangle$
are obtained through combined rotation and measurement operations. The diagonal elements
$\varrho_I = 1$ and $\varrho_Z = P_0-P_1$ come from direct
population measurement, while off-diagonal terms require pre-rotations:
\be
\varrho_X &=& \left\langle Y(-\pi/2)\psi\right. \left|Z\right|\left. Y(-\pi/2)\psi\right\rangle, \\
\varrho_Y &=& \left\langle X(\pi/2)\psi\right. \left|Z\right|\left. X(\pi/2)\psi\right\rangle.
\label{eq_s21}
\ee
implemented via additional gates $Y(-\pi/2)$ and $X(\pi/2)$ before measurement.

This method extends to $N$-qubit systems through the expansion of density matrix
$\rho=|\psi\rangle\langle\psi|$
over the Pauli operators as
$\rho=\sum_{\alpha}\varrho_{\alpha}\mathcal{O}_{\alpha}$,
where $\mathcal O_{\alpha}=\sigma_1\sigma_2\cdots\sigma_N$ is
a product of the Pauli operators
with $\sigma_{i=0,\cdots, N} \in \{I,X_i,Y_i,Z_i\}$.
Following the treatment in the 1-qubit
system, the extraction of $\varrho_\alpha$ requires the generation of a new quantum state,
\be
|\psi^\alpha\rangle = \mathcal A_{\sigma_1}\mathcal A_{\sigma_2}\cdots \mathcal A_{\sigma_N} |\psi\rangle.
\label{eq_s23}
\ee
For the Pauli operator $\sigma_i$ of $i$th qubit, the operator $\mathcal A_{\sigma_i}$
is given by
\be
\mathcal A_{\sigma_i} = \left\{ \ba{ll} I_i & ~~~~\sigma_i=I_i~\mathrm{and}~Z_i \\
Y_i(-\pi/2) & ~~~~\sigma_i= X_i \\
X_i(\pi/2) & ~~~~\sigma_i = Y_i
\ea.
\right.
\ee
Then, the population readout over the state $|\psi^\alpha\rangle$ can lead to the expansion
coefficient $\varrho_\alpha$.

Given the experimental overhead of full QST, we primarily use single-shot
population readout for VAE states, leveraging the real-valued coefficients
of silicon's Bloch states and determining signs via comparison with
numerical simulations.

\end{document}